\documentclass[aps,prf,longbibliography,preprint,superscriptaddress,showpacs,floatfix]{revtex4-1}
\usepackage{dcolumn}
\usepackage{bm}

\usepackage{colordvi,graphicx,color,amsbsy}
\usepackage{amsmath}
\usepackage{amssymb}
\usepackage[normalem]{ulem} %this works
\usepackage{comment}
\usepackage{appendix}

\begin{document}

%\preprint{}
\preprint{Submitted to Physical Review Fluids}

\title{Lagrangian and Eulerian Accelerations in Turbulent Stratified Shear Flows}


%\author{}
%\thanks{}
%\affiliation{}

\author{Frank G. Jacobitz}
 \email{jacobitz@sandiego.edu}
 \affiliation{Mechanical Engineering Department,
              Shiley-Marcos School of Engineering,
              University of San Diego,
              5998 Alcal\'a Park, San Diego, CA 92110, USA}
%\affiliation{Mechanical Engineering Program, University of San Diego, San Diego, CA 92110, USA}
%\affiliation{M2P2--CNRS \& CMI, Universit\'e de Provence, Marseille, France}
%\affiliation{Centre de Math\'ematiques et d'Informatique, Universit\'e de Provence,
%             39 rue Joliot-Curie, 13453 Marseille Cedex 13, France}

\author{Kai Schneider}
 \email{kai.schneider@univ-amu.fr}
 \affiliation{Aix-Marseille Universit\'e, CNRS, Centrale Marseille, Institut de Math\'ematiques de Marseille (I2M),
              39 rue Joliot-Curie, 13453 Marseille Cedex 13, France}
%Institut de Math\'ematiques de Marseille (I2M)
%Aix Marseille Univ, CNRS, Centrale Marseille, I2M, Marseille, France

%\author{Wouter J. T. Bos}
% \email{wouter.bos@ec-lyon.fr}
% \affiliation{LMFA--CNRS, Ecole Centrale de Lyon--Universit\'e de Lyon,
%              36 Avenue Guy de Collongue, 69134 Ecully Cedex, France}
%\affiliation{Laboratoire de M\'ecanique des Fluides et d'Acoustique
%             du Centre National de la Recherche Scientifique,
%             Ecole Centrale de Lyon, Universit\'e de Lyon,
%             69134 Ecully Cedex, France}

%\author{Marie Farge}
% \email{marie.farge@ens.fr}
% \affiliation{LMD--IPSL--CNRS, Ecole Normale Sup\'erieure,
%              24 rue Lhomond, 75231 Paris Cedex 5, France}
%\affiliation{Laboratoire de M\'et\'eorologie Dynamique du
%             Centre National de la Recherche Scientifique,
%             Ecole Normale Sup\'erieure,
%             24 rue Lhomond, 75231 Paris Cedex 5, France}


\date{\today}

\begin{abstract}
The Lagrangian and Eulerian acceleration properties of fluid particles in homogeneous turbulence with uniform shear and uniform stable stratification are studied using direct numerical simulations. 
The Richardson number is varied from $Ri=0$, corresponding to unstratified shear flow, to $Ri=1$, corresponding to strongly stratified shear flow. The probability density functions (pdfs) of both Lagrangian and Eulerian accelerations have a stretched-exponential shape and they show a strong and similar
influence on the Richardson number. 
The extreme values of the Eulerian acceleration are stronger than those observed for the Lagrangian acceleration. 
Geometrical statistics explain that the magnitude of the Eulerian acceleration is larger than its Lagrangian counterpart due to the mutual cancellation of the Eulerian and convective acceleration, as both vectors statistically show an anti-parallel preference.
A wavelet-based scale-dependent decomposition of the Lagrangian and Eulerian accelerations is performed. The tails of the acceleration pdfs grow
heavier for smaller scales of turbulent motion. Hence the flatness increases with decreasing scale, indicating stronger intermittency at smaller scales. 
The joint pdfs of the Lagrangian and Eulerian accelerations indicate a trend to stronger correlations with increasing Richardson number and
at larger scales of the turbulent motion. 
A consideration of the terms in the Navier--Stokes equation shows that the Lagrangian acceleration is mainly determined by the pressure-gradient
term, while the Eulerian acceleration is dominated by the nonlinear convection term. 
A similar analysis is performed for the Lagrangian and Eulerian time-rates of change of both fluctuating density and vorticity.
The Eulerian time-rates of change are observed to have substantially larger extreme values than those of their Lagrangian counterparts due to the the advection terms in the advection-diffusion equation for fluctuating density and in the vorticity equation, respectively. 
The Lagrangian time-rate of change of fluctuating vorticity is mainly determined by the vortex stretching and tilting term in the vorticity equation.
Since the advection-diffusion equation for fluctuating density lacks a quadratic term, the Lagrangian time-rate of change pdfs of fluctuating density show a more Gaussian shape, in particular for large Richardson numbers. 
Hence, the Lagrangian acceleration and time-rates of change of fluctuating density and vorticity reflect the dominant physics of the underlying
governing equations, while the Eulerian acceleration and time-rates of change are mainly determined by advection.
\end{abstract}

%\keywords{}

\maketitle

%\tableofcontents

%\section{}
%\subsection{}
%\subsubsection{}

% If in two-column mode, this environment will change to single-column
% format so that long equations can be displayed. Use
%\begin{widetext}
%\end{widetext}

\section{Introduction}
\label{sec:introduction}

An understanding of the Lagrangian acceleration properties of a fluid particle in turbulent motion is of fundamental importance and numerous applications exist in geophysical, environmental, and engineering flows. 
It aids in the study of transport and mixing, as well as in the characterization of geometric properties and intermittency at various scales of turbulent motion. 
As proposed in %Tsinober 
\citet{tsinvy2001} and \citet{tsinober2001}, the Lagrangian description of turbulent flows may be a more natural approach to the study of turbulence, as it is more directly related to the dynamics of fluid particles, which are subjected to different forces, e.g., due to the fluctuating pressure gradient, buoyancy, viscous stresses, or other forcing terms.

Studying Lagrangian acceleration has some history. After early work by \citet{heisenberg1948} and \citet{yaglom1949}, more recent studies range from theoretical investigations (e.g.~\citet{tsinvy2001, tsinober2001}) to applications such as the modeling of particle dispersion (e.g.~\citet{pope1994}) highly relevant to turbulent combustion. 
Such studies are carried out using both experimental (e.g.~\citet{laporta2001}) as well as computational (e.g.~\citet{yeung1989, yeung2002} or \citet{toschi2009}) approaches.

% previous work HIT
The majority of previous investigations focused on Lagrangian properties of isotropic turbulence. The
Lagrangian acceleration was found to be strongly intermittent and heavy tails were observed in its  probability density functions (pdf).
For example, extreme values as high as 1,500 times the acceleration of gravity were observed for the
Lagrangian acceleration of fluid particles by \citet{laporta2001} and numerical simulations by
\citet{toschi2009} confirmed these results. 

Acceleration fluctuations and the different contributions have been studied in \citet{pinsky2000,tsinvy2001} 
in isotropic turbulence. Their work is motivated by the random Taylor hypothesis or sweeping decorrelation 
hypothesis stating that `small eddies in turbulent flow being swept past a stationary Eulerian observer' 
\cite{tsinvy2001}.  It is based on the prediction of \citet{tennekes1975} that states that the Lagrangian 
acceleration must be small, justified by considering Eulerian and Lagrangian time scales. He predicted 
that the rms value of the Lagrangian acceleration is a factor $Re_{\lambda}^{-1/2}$ smaller than the Eulerian value.
\citet{lin1953} showed that there is no general justification to extend Taylor's hypothesis to 
turbulent shear flow. He gives some perspectives that this may still hold for large wavenumbers (small scales), 
which will be the topic of future work.

In \citet{tsinvy2001} direct numerical simulation data of isotropic turbulence were analyzed for different Reynolds numbers and 
the Lagrangian acceleration, called total acceleration in Tsinober's work, was decomposed into the Eulerian acceleration
(called local acceleration in \cite{tsinvy2001}) and the convective contribution. Possible cancellation 
properties between the Eulerian and convective contributions may yield reduced values of the Lagrangian
acceleration. The authors found that the variance of Lagrangian acceleration is much smaller than the Eulerian 
(local) acceleration and the advection term due to their strong negative alignment (or correlation) for 
sufficiently high $R_\lambda$, here 140. They also observed that the Lagrangian acceleration is strongly 
correlated with the pressure gradient. Their results are thus in support with the random Taylor hypothesis.

Note that the convective contribution of the acceleration becomes large when the flow is 
non-uniform, i.e., if the velocity changes along a streamline. The  convective  acceleration term is
nonlinear, which causes mathematical difficulties in flow analysis; also, even in steady flow (which is perfect for 
Taylor's hypothesis), the convective acceleration can be large if spatial gradients of velocity are large. 
In case it is anti-aligned with the local acceleration, it can be balanced and the total acceleration can 
still be small. This implies that the rate of Eulerian decorrelation is higher than that of Lagrangian 
decorrelation, which is crucial for two-point closures, see also the review on space-time correlations in turbulence by~\citet{he2017}.

Many applications of Lagrangian dynamics target the transport and mixing of natural and anthropogenic
substances in the geophysical environment. Such flows are often characterized by the presence of shear
and stratification. Homogeneous turbulent stratified shear flows with constant vertical stratification
rate $S_\rho=\partial \varrho/ \partial y$ and constant vertical shear rate $S=\partial U/ \partial y$
represents the simplest flow configuration in order to study the competing effects of shear and
stratification. This flow has been investigated extensively in the past. Experimental studies include
work by \citet{komori1983}, \citet{rohr1988}, \citet{piccirillo1997}, and \citet{keller2000}. Numerical
simulations were performed by \citet{gerz1989}, \citet{holt1992}, \citet{jacobitz1997}, \citet{jacobitz2002},
and \citet{portwood2019}. \citet{hanazaki2004} analyzed this flow using linear theory. More recently,
the mixing properties of turbulent stratified shear flow have been considered by, for example,
\citet{salehipour2015} and \citet{venayagamoorthy2016}. For a review, we refer to~\citet{gregg2018}.

More recently, \citet{jacobitz2016} considered Lagrangian
and Eulerian accelerations in rotating and sheared homogeneous turbulence. It was found that the
Lagrangian acceleration was mainly determined by the pressure-gradient term in the Navier--Stokes
equation, while the Eulerian acceleration shows stronger tails due to the advection term. In the
case of strong rotation, linear effects are dominant and the Lagrangian acceleration pdf takes an
approximately Gaussian shape. A comparison of linear theory with direct numerical simulation of
rotating and sheared homogeneous turbulence was performed by \citet{salhi2014}.
%kai: we could add a comment on the main results. ' nonlinear terms always reamin significant

The goal of this work is to investigate the acceleration statistics and to analyze the different 
contributions to the acceleration in turbulent stratified shear flows using direct numerical simulations. 
A key question is the understanding of the properties of Lagrangian acceleration fluctuations 
and their Eulerian counterpart and the influence of the Richardson number. 

In the following, the numerical approach taken in this study is introduced first. Then the Richardson number 
dependence of the Lagrangian and Eulerian acceleration pdfs are presented and geometrical statistics 
of the alignment angles of the different contributions. Using a wavelet-based scale-dependent decomposition, the 
Lagrangian and Eulerian accelerations are studied at various scales of the turbulent motion and their spatial 
fluctuations are analyzed. The corresponding Lagrangian and Eulerian time-rates of change pdfs for the 
%fluctuating vorticity and 
fluctuating density are discussed. Finally, a summary and conclusion of the present work is provided.
Results for the Lagrangian and Eulerian time-rates of change for the fluctuating vorticity as well as the Lagrangian and Eulerian acceleration component are discussed in the appendix.

\section{Approach}
\label{sec:approach}

In this section, the equations of motion and their direct numerical solution are described, variance estimates for the Lagrangian and Eulerian accelerations are given, the wavelet-based scale-dependent decomposition of the accelerations is introduced, and geometrical statistics to study the alignment of the different acceleration contributions are motivated.

\subsection{Equations of Motion}

The mean flow with velocity $(U, V, W)$ and density $\varrho$ considered in this study has a constant
vertical shear rate $S=\partial U/ \partial y$ and a constant vertical stratification rate
$S_{\rho}=\partial \varrho/ \partial y$, respectively:
\begin{equation}
  U = S y, \qquad V = W = 0, \qquad \varrho = \rho_0 + S_\rho y,
\end{equation}
where $\rho_0$ is the ambient density.

This study is based on the incompressible Navier--Stokes equations for the fluctuating velocity and an
advection-diffusion equation for the fluctuating density:
\begin{equation}
  \nabla \cdot \bm u = 0
\label{eqn_continuity}
\end{equation}
\begin{eqnarray}
\nonumber
  \frac{\partial \bm u}{\partial t}
  + \bm u \cdot \nabla \bm u
  + S y \frac{\partial \bm u}{\partial x}
  + S v \bm{e}_x \\
  = -\frac{1}{\rho_0} \nabla p
  -\frac{g}{\rho_0} \rho \bm{e}_y
  + \nu \nabla^2 \bm u
\label{eq:momentum}
\end{eqnarray}
\begin{equation}
  \frac{\partial \rho}{\partial t}
  + \bm u \cdot \nabla \rho
  + S_\rho v
  = \alpha \nabla^2 \rho
\label{eqn_density}
\end{equation}
Here, $\bm u = (u, v, w)$ is the fluctuating velocity, $p$ the fluctuating pressure, $\rho$ the
fluctuating density, $\nu$ the kinematic viscosity, and $\alpha$ the scalar diffusivity. Taking
the curl of the momentum  equation (\ref{eq:momentum}) leads to the vorticity equation:
\begin{eqnarray}
\nonumber
  \frac{\partial \bm \omega}{\partial t}
  + \bm u \cdot \nabla \bm \omega
  + \nabla \times \left( S y \frac{\partial \bm u}{\partial x} + S v \bm{e}_x \right) \\
  = \bm \omega \cdot \nabla \bm u
  - \nabla \times \left( \frac{g}{\rho_0} \rho \bm{e}_y \right)
  + \nu \nabla^2 \bm \omega
\end{eqnarray}

\subsection{Numerical Approach}

For their numerical solution, the equations of motion (\ref{eqn_continuity}-\ref{eqn_density}) are transformed into a frame of reference moving with the mean velocity (see \citet{rogallo1981}). 
This transformation enables the application of periodic boundary conditions for the fluctuating components of velocity and density. 
A spectral collocation method is used for the spatial discretization and the solution is advanced in time with a fourth-order Runge--Kutta scheme.

Table \ref{tab_param} provides an overview of the simulations performed for this study.
The Richardson number $Ri=N^2/S^2$ is varied from $Ri=0$, corresponding to unstratified shear flow, to $Ri=1$, corresponding to strongly stratified shear flow. 
While both the mean shear rate $S=\partial U/ \partial y$ and the mean stratification rate $S_{\rho}= \partial \varrho/ \partial y$ are constant for a given simulation, the Richardson number variation is obtained by a change of the Brunt--V\"ais\"al\"a frequency $N$ with $N^2=-g/\rho_0 S_{\rho}$, while keeping the mean shear rate $S$ constant.

The initial conditions are taken from a separate simulation of isotropic
turbulence without density fluctuations, which was allowed to develop for approximately one eddy turnover time. 
The initial values of the Taylor-microscale Reynolds number $Re_\lambda q \lambda/\nu = 89$ and the shear number $SK/\epsilon = 2$ are fixed. 
Here $q$ is the rms of the fluctuating velocity with $q^2 = \overline{u_i u_i}$, $\lambda$ the Taylor-microscale with $\lambda^2 = 5 q^2 \nu/\epsilon$, $K=q^2/2$ the kinetic energy, and $\epsilon = \nu
\overline{\partial u_j/\partial x_k \partial u_j/\partial x_k}$ the dissipation rate. 

Table \ref{tab_param} provides an overview of the eventual values of $Re_\lambda$, $q$, and $\epsilon$ at time $St=10$. The table also lists values of a variety of length scales, including the overturning scale $L_{overturn} = q^3/\epsilon$, the Ellison scale $L_{Ellison} = \rho/S_\rho$, the Ozmidov scale $L_{Ozmidov}$ with $L_{Ozmidov}^2 = \epsilon/N^3$, the Taylor-microscale $\lambda$, and the Kolmogorov scale $\eta$ with $\eta^4 = \nu^3/\epsilon$, indicating an appropriate resolution of the simulations at $St=10$ at both the large and small scales of the turbulent motion.

The simulations are performed on a parallel computer using $512 \times 512 \times 512$ grid points.
To increase the resolution, instead of the classical dealiasing with a cut-off at $2/3$ of the maximum wavenumber, a cosine-filter dealiasing is applied.
The cosine-filter is only applied to wavenumbers larger than $2/3$ of the maximum wavenumber. Its transfer function starts with one at $2/3$ of the maximum wavenumber, goes to zero at the maximum wavenumber, and it follows the shape of the first quarter of the cosine function period.
The maximum wavenumber $k_{max}$ can be defined when the cosine is equal to the value $1/2$.
For the current resolution with $N= 512$ we thus have $k_{max} = 227$ (instead of the value 170 obtained for classical dealiasing).
All simulations are well resolved and we have $k_{max} \eta > 1.2$ in the eventual evolution for the unstratified case.
A discussion on the influence of dealising in pseudo-spectral codes can be found in \citet{hou2007}. The authors show that the classical $2/3$ rule  does not necessarily yield the best results and other filtering techniques, different from the cosine-dealiasing used here, are more efficient, supporting our choice.

\subsection{Variance Estimates for the Lagrangian and Eulerian Accelerations in Stratified Shear Flow}

The Lagrangian and Eulerian accelerations are defined as
\begin{equation}
\bm a_L=\frac{\partial \bm u}{\partial t} + \bm u \cdot \nabla \bm u
\quad
\textrm{and}
\quad
\bm a_E=\frac{\partial \bm u}{\partial t},
\end{equation}
respectively. Both accelerations are computed as a volume average at a fixed time, which is an appropriate
choice for homogeneous flows. The effects of shear and buoyancy are considered as external forces.

In \cite{jacobitz2016} we provided estimates of the variances of the Lagrangian and Eulerian accelerations
writing the Navier--Stokes equations in the form
\begin{equation}
  \frac{\partial \bm u}{\partial t} = -\bm N -\bm \Pi -\bm \Lambda,
\label{eq:NSEI}
\end{equation}
where the terms on the right hand side are given by
\begin{eqnarray}
        \bm N & = & \bm a_C = \bm u \cdot \nabla \bm u \nonumber \\
      \bm \Pi & = & \bm a_P = \nabla (p/\rho_0) \nonumber \\
  \bm \Lambda & = & \bm \Lambda_S + \bm \Lambda_B + \bm \Lambda_V = S v \bm e_x + \frac{g}{\rho_0} \rho \bm e_y - \nu \nabla^2 \bm u.
\label{eq:Li}
\end{eqnarray}
Here, $\bm N$ is the nonlinear or advection term, $\bm \Pi$ the pressure-gradient term, and $\bm \Lambda$ the linear term with
contributions from shear, buoyancy, and viscous effects. The notation 
$\bm a_C$ for the nonlinear term and $\bm a_P$ for the pressure gradient
match the notation in \citet{tsinvy2001} to denote the convective and pressure contributions, respectively, to the accelerations.

According to \cite{jacobitz2016} we have also in the case of stratified shear flow the identity
\begin{equation}
  \left<\|\bm N + \bm \Pi +\bm \Lambda\|^2 \right> = \left<\|\bm N + \bm \Lambda \|^2 \right> - \left<\|\bm \Pi\|^2 \right>.
\label{eq:Id}
\end{equation}
Here, $\| \cdot \|$ denotes the magnitude of a vector and $\left< \cdot \right>$ the volume average for a homogeneous field.

This directly implies the following exact identities 
for the variances of the Eulerian acceleration ${\bm a}_E$ (called local acceleration in \citet{tsinvy2001}),
\begin{equation} 
\label{eqn_theory_vaae}
  a_E^2 \equiv \left<\|\frac{\partial \bm u}{\partial t}\|^2\right>
  = \left<\|\bm N + \bm \Pi +\bm \Lambda\|^2 \right>
  = \left<\|\bm N + \bm \Lambda\|^2\right> - \left<\|\bm \Pi\|^2\right>
\end{equation}
and of the Lagrangian acceleration ${\bm a}_L$ (called total acceleration in \citet{tsinvy2001}),
\begin{equation}
\label{eqn_theory_vaal}
  a_L^2 \equiv \left< \|\frac{\partial \bm u}{\partial t} + \bm u \cdot \nabla \bm u\|^2 \right>
  = \left<\|\bm \Pi +\bm \Lambda\|^2 \right>.
\end{equation}

The variance estimates provided in \cite{jacobitz2016} for rotating shear flows are now extended to
stratified shear flows, again with the underlying assumption and crucial simplification of isotropy
of the flow. The main difference arises in the linear term $\bm \Lambda$, which now includes a
buoyancy force, instead of a Coriolis force. Neglecting the friction force, the variance of the linear
term can be written as
\begin{equation}
\Lambda^2 = \frac{1}{3} S^2 [1 + 3 (\frac{g}{\rho_0})^2 \frac{1}{S^2} \frac{\rho^2}{u^2}] u^2.
\end{equation}
Using the ratio of potential to kinetic energy
\begin{equation}
\frac{K_\rho}{K} = \frac{-\frac{1}{2} \frac{g}{\rho_0} \frac{\rho^2}{S_\rho}}{\frac{1}{2} u^2}
                 = -\frac{g}{\rho_0} \frac{1}{S_\rho} \frac{\rho^2}{u^2}
\end{equation}
the variance for the linear term can be written as
\begin{equation}
\label{eqn_theory_valin}
\Lambda^2 = \frac{1}{3} S^2 [1 + 3 Ri \frac{K_\rho}{K}] u^2.
\end{equation}
Hence, the variance estimate of the linear term retains the ratio of potential to kinetic energy.

\subsection{Scale-dependent decomposition of Lagrangian and Eulerian accelerations}

To gain insight into the scale dependence of the Lagrangian and Eulerian accelerations, we decompose
both accelerations into an orthogonal wavelet series. Wavelets are well localized functions in space and
in scale (or wave number), see e.g.~\cite{mallat1998}, and different wavelet-based diagnostics, including
the scale-dependent energy distribution and its spatial fluctuations, intermittency measures such as the
scale dependent flatness and anisotropy measures, have been proposed. For a review we refer the reader to
\cite{farge2015}.

We consider a generic vector field ${\bm a} = (a_1, a_2, a_3)$ at a fixed time instant and decompose
each component $a_\alpha ({\bm x})$ into an orthogonal wavelet series,
\begin{equation}
a_\alpha ({\bm x}) \, = \, \sum_{\lambda} \, \widetilde a^{\alpha}_{\lambda} \, \psi_{\lambda} ({\bm x}),
\label{ows}
\end{equation}
where the wavelet coefficients are given by the scalar product $ \widetilde a^{\alpha} = \langle a_{\alpha},
\psi_{\lambda} \rangle$. The wavelets $\psi_{\lambda}$ with the multi-index $\lambda = (j,{\bm i}, d)$ are
well localized in scale $L_0 2^{-j}$ (where $L_0$ corresponds to the size of the computational domain),
around position $L_0 {\bm i}/ 2^j$, and orientated in one of the seven directions $d= 1, ..., 7$, respectively.
The scale is directly related to the wave number $k_j = k_0 2^j$, where $k_0$ is the centroid wave number of
the chosen wavelet family. For Coiflets 12 wavelets used in the present work we have $k_0 = 0.77$. Large scales
correspond to small values of the scale index $j$ and to a well localized wavelet in Fourier space around
wavenumber $k_j$. In contrast for small scales, which correspond to large values of $j$, the wavelet becomes
less localized in Fourier space around the mean wavenumber $k_j$.

Reconstructing the three components $a^{\alpha}$ at scale $2^{-j}$ by summing only over the position ${\bm i}$
and direction $d$ indices in eq.~\ref{ows} yields the acceleration ${\bm a}^j$ at scale intex $j$. In terms of
filtering the acceleration at a given scale corresponds to a bandpass filtered field, with a bandpass filter
having constant relative bandwidth. This means that the filterwidth becomes larger at larger wavenumber,
corresponding to decreasing scale. By construction we have ${\bm a} = \sum_{j} {\bm a}^j$, where the ${\bm a}^j$
are mutually orthogonal.

The scale-dependent moments, including scale-dependent flatness, and scale-dependent pdfs, can thus be
computed from ${\bm a}^j$ using classical statistical estimators.

For instance the $q$-th order moment of ${\bm a}^j({\bm x})$ can be defined by,
\begin{equation}
M_q [ {\bm a}^j ] = \langle ( {\bm a}^j )^q \rangle, 
\end{equation}
and by construction the mean value vanishes, $\langle {\bm a}^j \rangle=0$. 
The moments are thus central moments.
These scale-dependent moments are directly related to the $q$-th order structure functions~\cite{Schneider2004} where the increment size is $\propto 2^{-j}$.

%Ref:
%\bibitem{SFK04} K.~Schneider, M.~Farge, and N.~Kevlahan, {\it Spatial intermittency in two-dimensional turbulence: a wavelet approach, in Woods Hole Mathematics, Perspectives in Mathematics and Physics}, edited by N.~Tongring and R.~C.~Penner (World Scientific, Singapore, 2004), Vol. 34, pp. 302--328.

The scale-dependent flatness, which measures the intermittency of ${\bm a}^j$ at scale $2^{-j}$, is defined by
\begin{equation}
Fl [{\bm a}^j ]=\frac{M_4 [{\bm a}^j]}{ \left(M_2 [ {\bm a}^j  ] \right)^2 }.\label{eq:scale-dep-flatness}
\end{equation}
For a Gaussian distribution the flatness equals three at all scales.

\subsection{Geometrical statistics}
\label{sec:geostat}

To understand the magnitude of Eulerian and Lagrangian accelerations, we statistically assess, following %Tsinober et al. 
\citet{tsinvy2001}, the alignment properties of ${\bm a}_E$, ${\bm a}_C =\bm N$ and its sum corresponding to the Lagrangian acceleration ${\bm a}_L = {\bm a}_E + {\bm a_C}$. For convenience,
we partly use the notation introduced in \cite{tsinvy2001} in this section.
When the vectors of the Eulerian acceleration ${\bm a}_E$ and the convective terms ${\bm a}_C$ are anti-parallel, then the magnitude of the Lagrangian acceleration ${\bm a}_L$, is small compared to those of the Eulerian and convective contribution, since
\begin{equation}
\langle {\bm a}_L {\bm a}_L \rangle =
 \langle {\bm a}_E + {\bm a}_C, {\bm a}_E + {\bm a}_C\rangle = \langle {\bm a}_E, {\bm a}_E \rangle + \langle {\bm a}_C, {\bm a}_C \rangle + 2 \cos({\bm a}_E,{\bm a}_C) \, ||{\bm a}_E|| \, || {\bm a}_C ||.      
\end{equation}

%Note (for us) $\langle a + b, a+b \rangle = ||a||^2 + ||b||^2 + 2 ||a|| ||b|| \cos(a,b)$. 
If ${\bm a}_E$ and ${\bm a}_C$ are anti-aligned the cosine is negative and the norm of ${\bm a}_E + {\bm a}_C$ is minimal.
To verify the random Taylor-hypothesis \citet{tsinvy2001} computed the cosine of the angle of the Eulerian acceleration and the convective term, motivated by the prediction of \citet{tennekes1975} that the Lagrangian acceleration must be small so that the hypothesis holds. 
We expect this result to hold with modification due to shear and stratification.

For the pressure gradient term ${\bm a}_P = \bm \Pi$, the alignment with the Eulerian and Lagrangian acceleration can be likewise assessed. %Why? ($a_L + \Pi \approx 0$) 
For sufficiently high Reynolds numbers we anticipate a strong anti-alignment of ${\bm a}_P$ with ${\bm a}_L$ showing that the flow is driven by the pressure gradients and that linear effects are negligible. However, buoyancy may change this result for strong stratification and its impact will be assessed using the simulation results.

For Gaussian divergence free random fields \citet{tsinvy2001} found similar alignment properties and they concluded that the cancellation of $\bm a_E$ and $\bm a_C$ is mostly a kinematic effect and not due to Navier--Stokes dynamics. A justification for these findings is given by Millionshchikow's zero-fourth cumulant hypothesis \citep{moninyaglom1971} which decomposes fourth order moments into a series of second order moments.

%Statistical Fluid Mechanics: The Mechanics of Turbulence, volume 1. By A. S. MONIN and A. M. YAGLOM. M. I. T. Press, 1971

%Future work, check alignment scale dependent for HIT, this should go into the conclusions.

%--------------------------------------------------------------------------------------------
\section{Results}
\label{sec:results}

In this section, the flow evolution is briefly described first. Then, results for the probability density functions (pdfs) of the Lagrangian
and Eulerian accelerations are provided, related to the remaining terms in the Navier--Stokes equations, and
their scale-dependent properties are presented.  A similar analysis is performed for the Lagrangian and Eulerian time-rates of change of fluctuating density and in appendix A also for fluctuating vorticity. 
In the following, the accelerations and time-rates of change are analyzed at the instant $St=10$. 
Table~\ref{tab_param} provides an overview of the series of five simulations performed.

\subsection{Turbulence Evolution}

% figures 1 and 2

In order to provide a context for the present study, the energetics of the flow is briefly discussed.
More details on turbulent stratified shear flows can be found in \cite{jacobitz1997} and \cite{jacobitz2002}.

\begin{figure*} % 01
\setlength{\unitlength}{0.45\textwidth}
\includegraphics[bb=50 50 302 251,width=1\unitlength,angle=0]{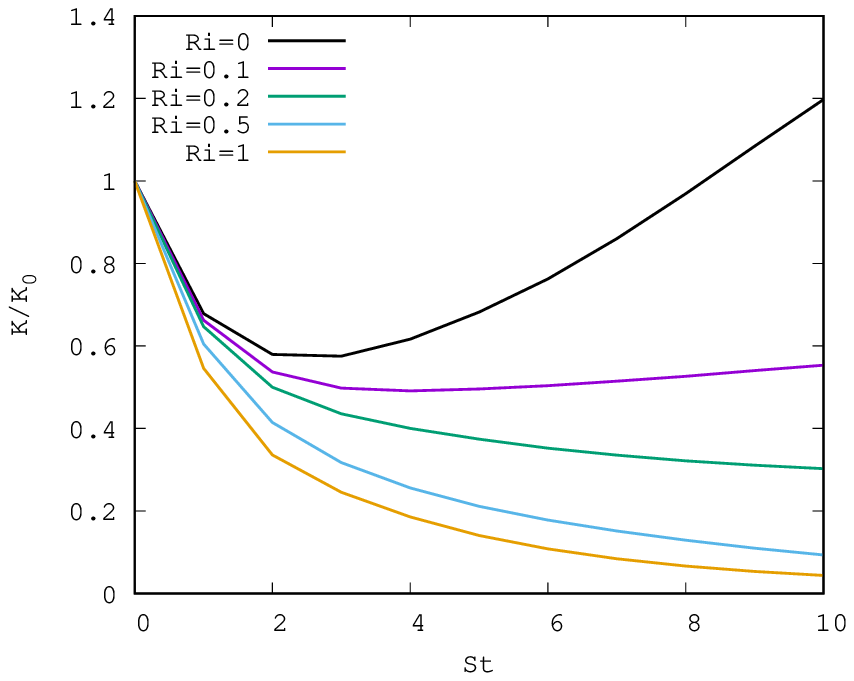}
\includegraphics[bb=50 50 302 251,width=1\unitlength,angle=0]{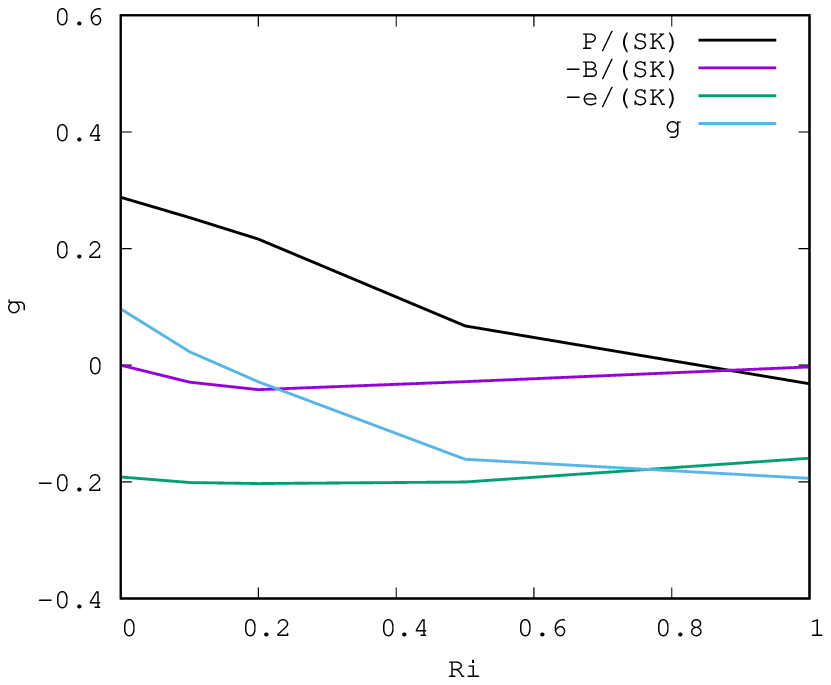}
\vspace{1.0cm}
\caption{Evolution of the turbulent kinetic energy $K$ in nondimensional time $St$ (left)
         and dependence of the normalized production rate $P/(SK)$, buoyancy flux $B/(SK)$,
         and dissipation rate $\epsilon/(SK)$ on the Richardson number at $St=10$ (right).}
\label{fig_kinetic}
\end{figure*}

Figure~\ref{fig_kinetic} (left) shows the evolution of the turbulent kinetic energy normalized by its
initial value $K/K_0$. All cases result in an initial decay phase due to the isotropic initial conditions.
Then, as the Richardson number $Ri$ is increased, the eventual evolution of the turbulent kinetic energy
changes from growth to decay with a critical value of $Ri_{cr} \approx 0.15$.

The normalized transport equation for the turbulent kinetic energy evolution can be written as:
\begin{equation}
\gamma = \frac{1}{SK} \frac{dK}{dt} = \frac{P}{SK} - \frac{B}{SK} - \frac{\epsilon}{SK}
\end{equation}
Here, $\gamma$ is the growth rate of the turbulent kinetic energy, $P/(SK)$ is the normalized production
term with $P = -S \overline{u_1 u_2}$, $B/(SK)$ is the normalized buoyancy flux with $B = g/\rho_0
\overline{u_2 \rho}$, and $\epsilon/(SK)$ is the normalized dissipation rate.

Figure~\ref{fig_kinetic} (right) shows the dependence of $P/(SK)$, $B/(SK)$, $\epsilon/(SK)$, and $\gamma$
on the Richardson number $Ri$ at nondimensional time $St = 10$. The normalized production rate $P/(SK)$
decreases with increasing Richardson number $Ri$ and it assumes a slightly negative value for large $Ri$
cases, indicating a positive Reynolds shear stress (or counter-gradient flux). The normalized buoyancy flux $B/(SK)$ remains relatively small
and it converts kinetic to potential energy for most of the $Ri$ range. The normalized dissipation rate
$\epsilon/(SK)$ remains relatively unaffected by the $Ri$ variation. The growth rate $\gamma$ follows the
trend of the normalized production rate $P/(SK)$, offset by the contributions of $B/(SK)$ and $\epsilon/(SK)$.
Note that positive values of $\gamma$ correspond to a growth of $K$, while a negative value of $\gamma$
indicates decay of the turbulent kinetic energy.

\begin{figure*} % 02
\setlength{\unitlength}{0.45\textwidth}
\includegraphics[bb=50 50 302 251,width=1\unitlength,angle=0]{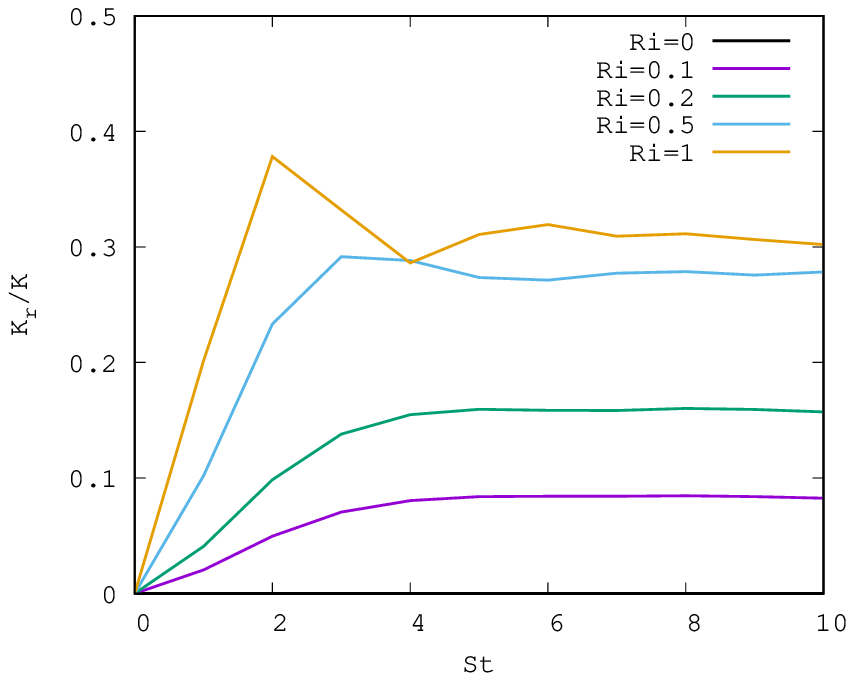}
\includegraphics[bb=50 50 302 251,width=1\unitlength,angle=0]{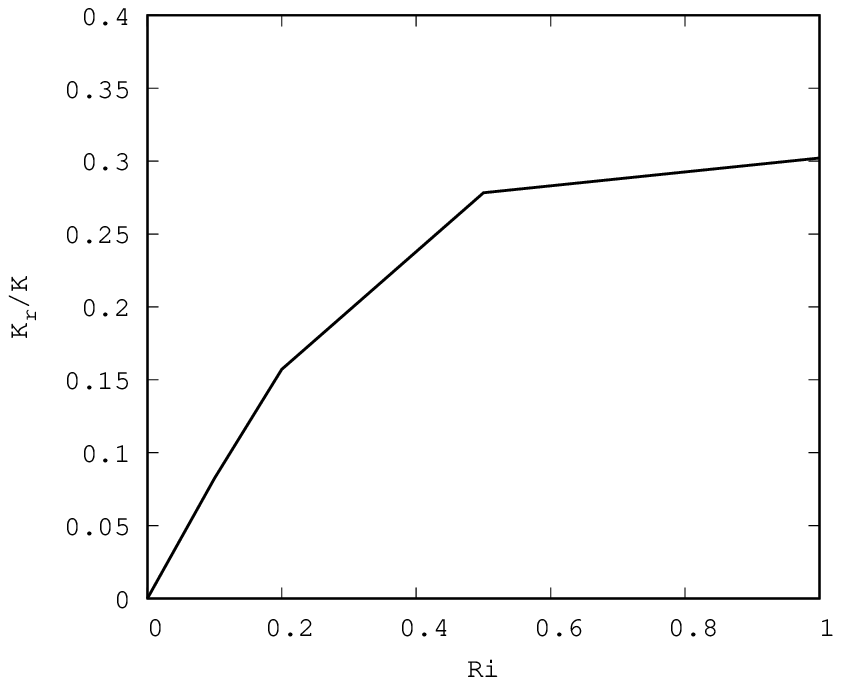}
\vspace{1.0cm}
\caption{Evolution of the ratio of turbulent potential to kinetic energy $K_{\rho}/K$ in nondimensional
         time $St$ (left) and dependence of this ratio on the Richardson number at $St=10$ (right).}
\label{fig_potential}
\end{figure*}

The evolution of the ratio of potential to kinetic energy is given in figure~\ref{fig_potential} (left).
The simulations are initialized without potential energy and a strong initial growth is observed. The
ratio of potential to kinetic energy eventually reaches an approximately constant value, which still
depends on the Richardson number $Ri$. This dependence of the ratio $K_\rho/K$ on $Ri$ at nondimensional
time $St = 10$ is presented in figure~\ref{fig_potential} (right). The ratio of $K_\rho/K$ first increases
strongly and then reaches a maximum of $K_\rho/K \approx 0.3$ for $Ri = 1$.
%, and finally slightly decreases for large $Ri$.

\subsection{Lagrangian and Eulerian Accelerations}

% figures 3 and 4

\begin{figure*} % 03
\setlength{\unitlength}{0.45\textwidth}
\includegraphics[bb=50 50 302 251,width=1\unitlength,angle=0]{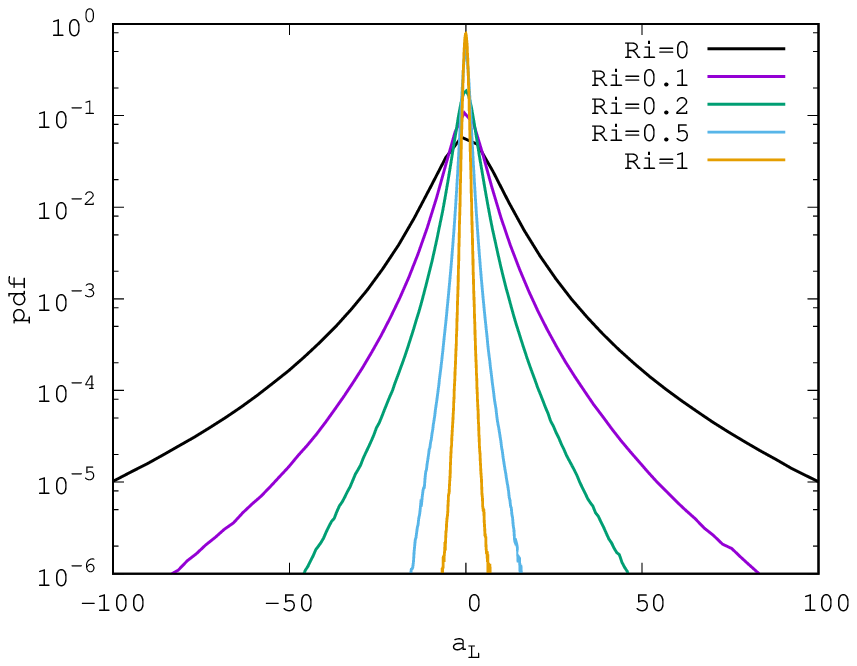}
\includegraphics[bb=50 50 302 251,width=1\unitlength,angle=0]{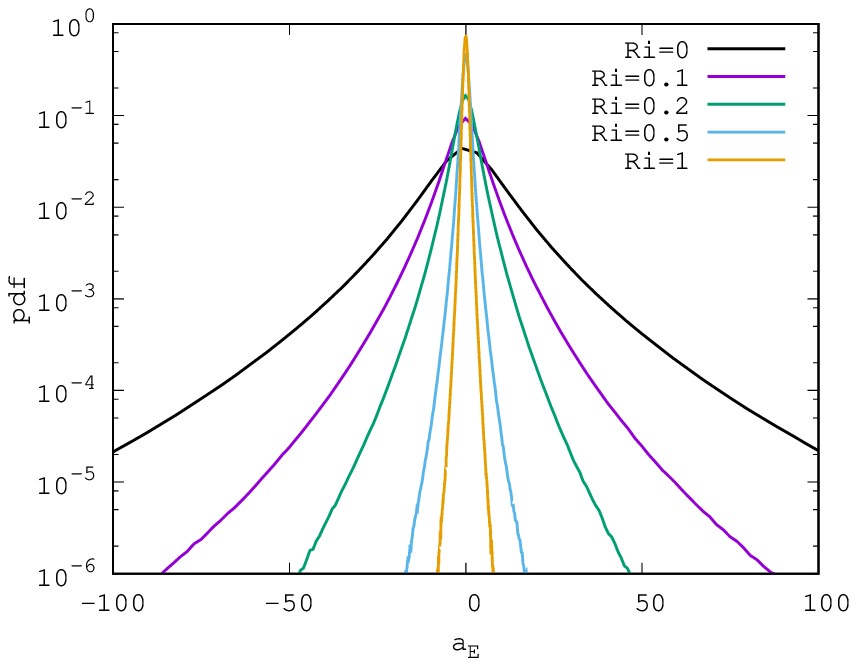}\\
\vspace{1.0cm}
\includegraphics[bb=50 50 302 251,width=1\unitlength,angle=0]{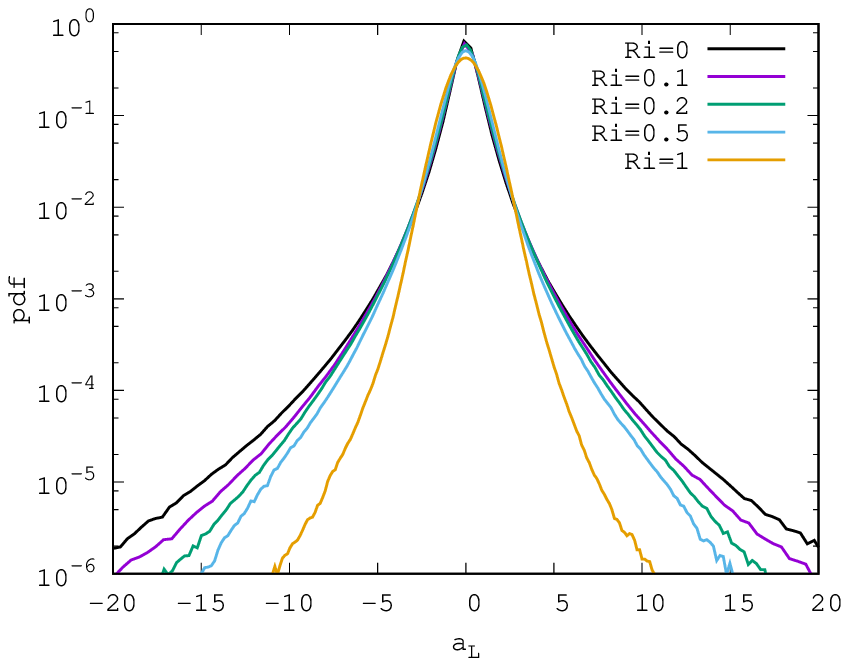}
\includegraphics[bb=50 50 302 251,width=1\unitlength,angle=0]{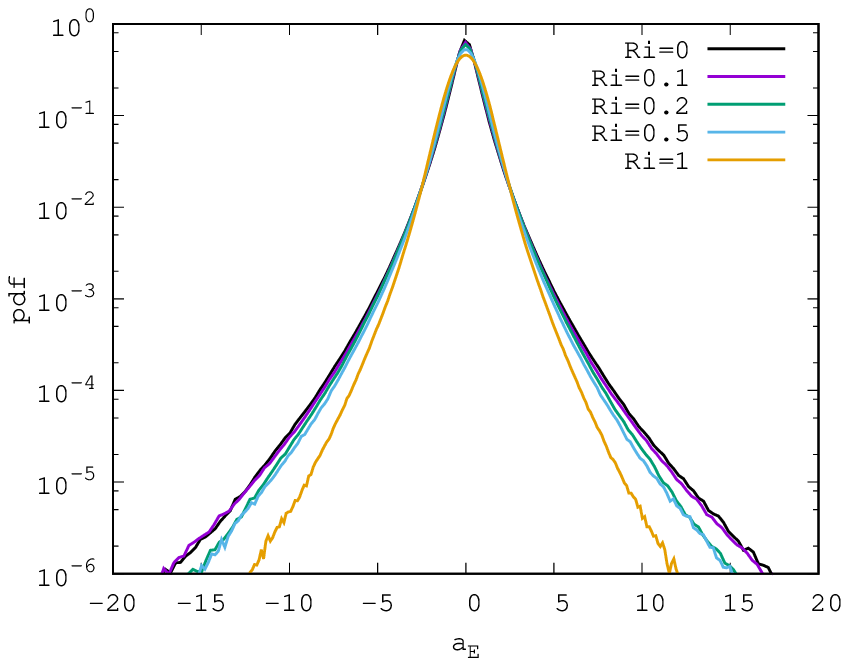}
\vspace{1.0cm}
\caption{Pdfs (top) and pdfs normalized with the corresponding standard deviations (bottom) of Lagrangian acceleration $\bm a_L$ (left) and 
         Eulerian acceleration $\bm a_E$ (right) at nondimensional time $St=10$. 
         Note that all pdfs are estimated using histograms with 100 bins, and they are plotted 
         in log-lin representation. Note that pdfs for the vector quantities are shown.}
\label{fig_vpdf+fig_vpdfn}
\end{figure*}

Figure \ref{fig_vpdf+fig_vpdfn} (top) shows the probability distribution functions (pdfs) of the Lagrangian acceleration
$\bm a_L$ (left) and of the Eulerian acceleration $\bm a_E$ (right). The pdfs of both accelerations have
stretched-exponential shapes and they exhibit a strong and similar influence on the Richardson number $Ri$.
For small $Ri$, the extreme values of the Eulerian acceleration are above those of the Lagrangian acceleration,
which is consistent with previous observations for sheared and rotating turbulence \citep{jacobitz2016} and likewise observed for isotropic turbulence by \citet{tsinvy2001}.

Figure \ref{fig_vpdf+fig_vpdfn} (bottom) shows the pdfs normalized with the corresponding standard deviations of the two accelerations. For a core region of about five
standard deviations, both the Lagrangian and Eulerian accelerations show approximately the same shape. The
tails of the pdfs of both accelerations weaken with increasing $Ri$. For small $Ri$, the tails of the
Lagrangian acceleration are heavier than the tails of the Eulerian acceleration.

% tabels 1 and 2

Tables \ref{tab_la} and \ref{tab_ea} provide statistical information of the Lagrangian and Eulerian
accelerations as a function of the Richardson number at nondimensional time $St = 10$. The magnitudes
(rms values) of both accelerations decrease with increasing $Ri$ and the magnitude of $\bm a_E$ remains
always larger than the magnitude of $\bm a_L$, also observed for isotropic turbulence \citet{tsinvy2001}. 
At small $Ri$, the heavier tails observed for the normalized
pdfs of $\bm a_L$ as compared to $\bm a_E$ result in a larger flatness of the Lagrangian acceleration pdf
as compared to its Eulerian counterpart. The flatness values of both accelerations generally decrease with
increasing $Ri$, indicating a decreased importance of nonlinear effects which is related to the decreasing Reynolds number. However, unlike in the presence
of strong rotation considered in \cite{jacobitz2016}, the flatness values do not reach a value close to
three, characteristic for a Gaussian distribution, in the case of strong stratification. 
For $Ri=0$ the flatness values of $\bm a_L$ and $\bm a_E$ are comparable with DNS data of isotropic turbulence \cite{tsinober2001} at $Re_{\lambda} =140$, where the values of $24.4$ and $12.8$ were found, while we find respectively 27.81 and 14.41 in the case of pure shear.
Note that for the variances \citet{tsinvy2001} found the values (normalized with $\epsilon^{3/2} \nu^{-1/2} = 24.84$, where $\epsilon = 1.20$ and $\nu = 0.0028$) of $2.75$ for $a_L$ %(non normalized $68.32$) 
and $8.19$ for $a_E$  %$(203.46)$ 
(and 10.96 for $a_C$), while we find respectively the values  
8.20 for $a_L$ %19.43  
and 14.81 for $a_E$ %26.10
(and 22.39 for $a_C$)
in the case of pure shear using the same normalization with $\epsilon^{3/2} \nu^{-1/2}$. This shows that the order is consistent and the values are comparable.

% figure 4

\begin{figure*} % 04
\setlength{\unitlength}{0.45\textwidth}
\includegraphics[bb=50 50 302 251,width=1\unitlength,angle=0]{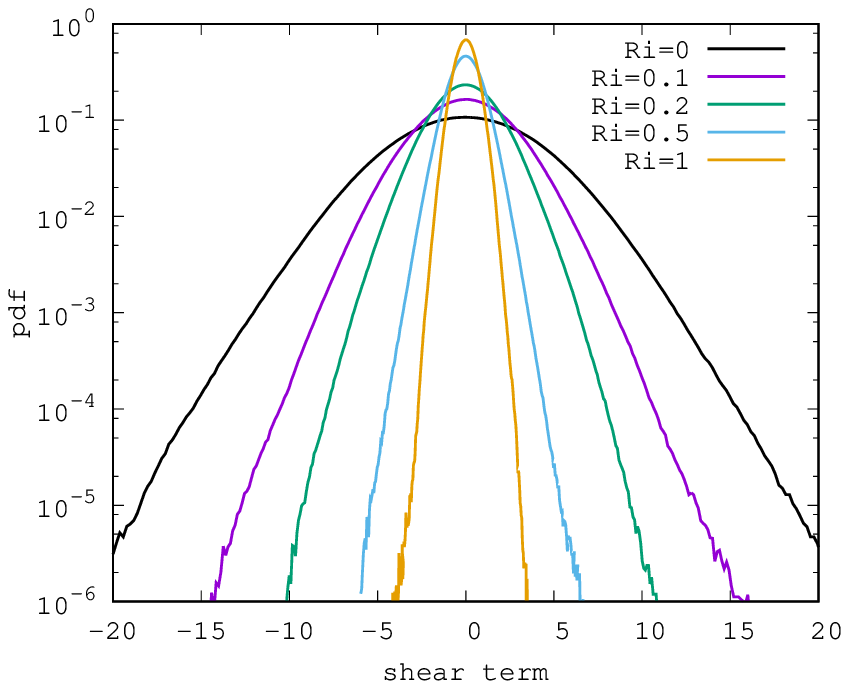}
\includegraphics[bb=50 50 302 251,width=1\unitlength,angle=0]{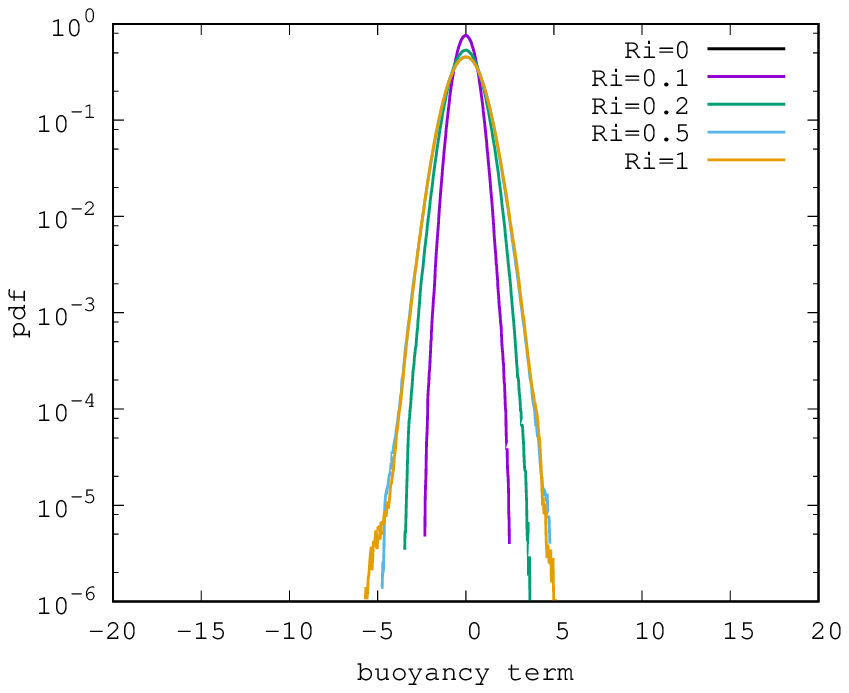}\\
\vspace{0.5cm}
\includegraphics[bb=50 50 302 251,width=1\unitlength,angle=0]{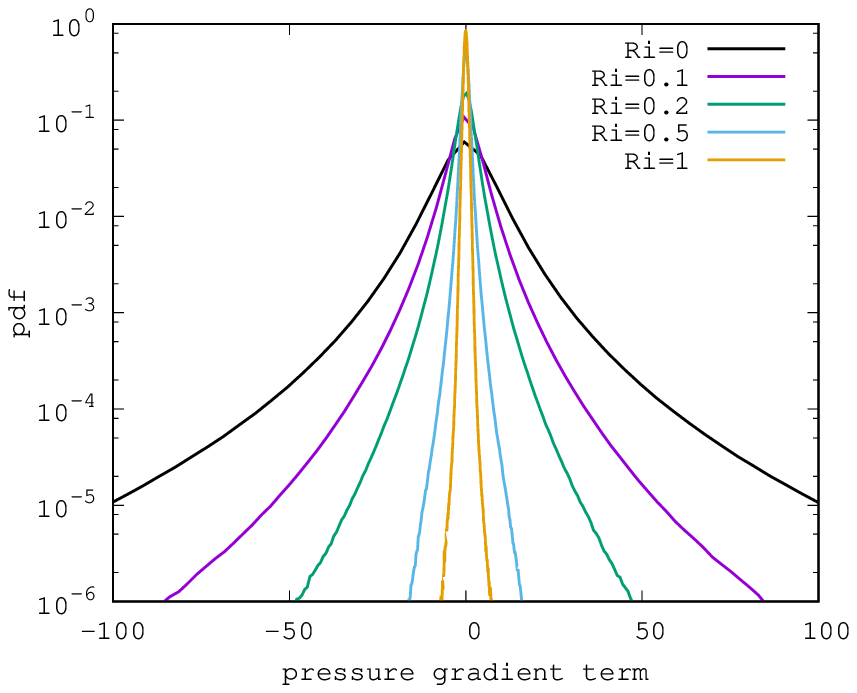}
\includegraphics[bb=50 50 302 251,width=1\unitlength,angle=0]{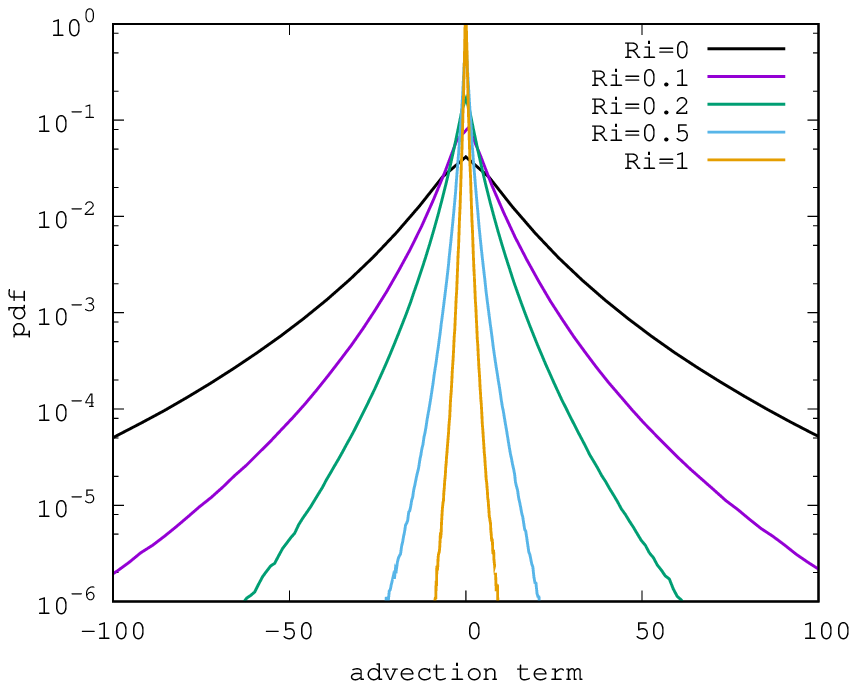}\\
\vspace{1.0cm}
\caption{Pdfs of the shear (top, left), buoyancy (top, right), pressure-gradient (bottom, left),
         and advection (bottom, right) terms in the Navier--Stokes equations at nondimensional time
         $St=10$.}
\label{fig_nse}
\end{figure*}

Figure \ref{fig_nse} shows pdfs of the shear term (top, left), the buoyancy term (top, right), the
pressure-gradient term (bottom, left), and the advection term (bottom, right) in the Navier--Stokes
equation. The shear and buoyancy terms depend linearly on fluctuating velocity components and density
and their pdfs have hence a Gaussian shape. While the magnitude of the shear term pdf decreases with
increasing $Ri$, the magnitude of the buoyancy term pdf increases. The pdfs of the pressure-gradient
and advection terms show a stretched-exponential shape due to the quadratic nature of the terms. The
magnitudes of both terms decrease with increasing $Ri$. For small $Ri$, the pressure-gradient and
advection terms clearly dominate the shear and buoyancy terms, but this dominance somewhat diminishes
with increasing $Ri$. Hence, the pressure-gradient term is the generally dominant contribution to the
Lagrangian acceleration, while the advection term is important for the Eulerian acceleration.

% table 3

Table \ref{tab_est} provides the variances of the contributions to the linear term from the shear term
$\Lambda_S^2$, the buoyancy term $\Lambda_B^2$, and the viscous term $\Lambda_V^2$. An estimate for the
variance of the linear term $\Lambda_{DNS}^2$ is computed using the triangle inequality. The table also
provides the variance of the velocity $q^2$, the ratio of potential to kinetic energies $K_\rho/K$, and
a theoretical estimate for the linear term $\Lambda^2$ based on equation \ref{eqn_theory_valin} at
nondimensional time $St=10$. The variances of the linear term computed from the simulation results
$\Lambda_{DNS}^2$ and the theoretical estimate $\Lambda^2$ agree well, despite the assumption of
isotropy used in the derivation of equation \ref{eqn_theory_valin}. The variance of the linear term
decreases with increasing Richardson number $Ri$.
The Taylor-microscale Reynolds number $Re_\lambda$ given in table~\ref{tab_param} yields
a measure for the general importance of nonlinear effects in a turbulent flow. $Re_\lambda$
decreases with increasing $Ri$.
The results suggest that nonlinear effects contribute the least to
the turbulence evolution for the case with a Richardson number $Ri=1$.

% figure 7

\begin{figure*} % 05
\setlength{\unitlength}{0.45\textwidth}
\includegraphics[bb=50 50 302 251,width=1\unitlength,angle=0]{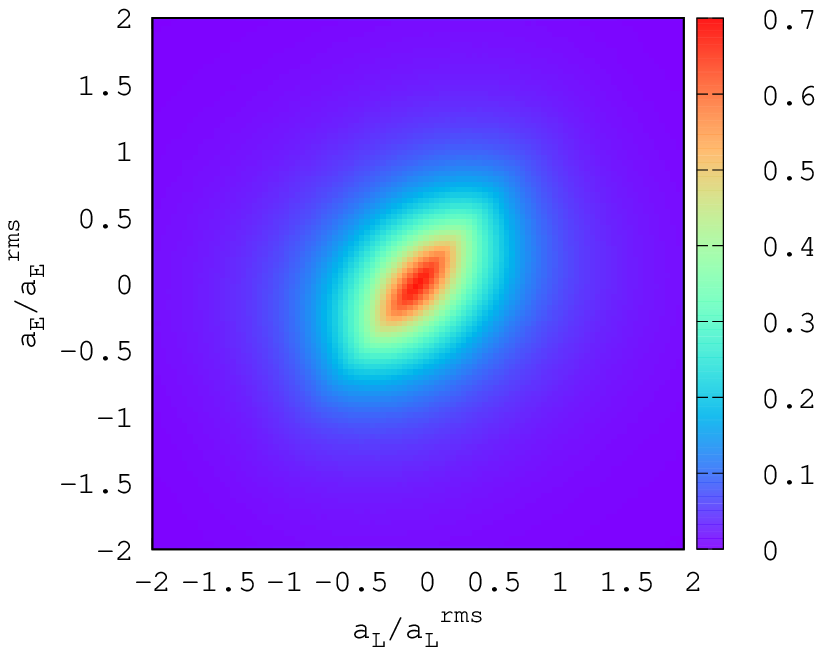}
\includegraphics[bb=50 50 302 251,width=1\unitlength,angle=0]{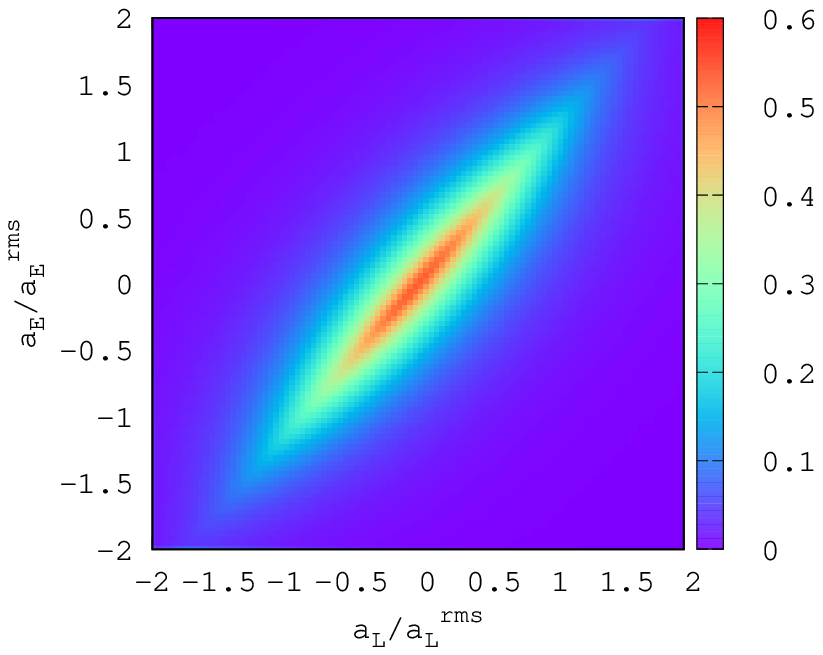}
\vspace{1.0cm}
\caption{Joint pdfs of Lagrangian acceleration $\bm a_L$ and Eulerian acceleration $\bm a_E$ for
         Richardson numbers $Ri=0.1$ (left) and $Ri=1$ (right) at nondimensional time $St=10$ using a linear color scale.}
\label{fig_jpdf}
\end{figure*}

The joint pdfs of the Lagrangian and Eulerian accelerations are shown in figure \ref{fig_jpdf} for
two cases with Richardson numbers $Ri=0.1$ (left) and $Ri=1$ (right) at nondimensional time $St=10$.
The correlation between Lagrangian and Eulerian accelerations is observed to increase with increasing
$Ri$. The stronger correlation of the Eulerian and Lagrangian acceleration for $Ri=1$ is due to the reduced nonlinearity. 

% table 4

In order to quantify this observation, the Pearson product-moment correlation coefficient for the
Lagrangian and Eulerian accelerations in dependence of the Richardson number $Ri$ is given in the
first line of table \ref{tab_geostat} at nondimensional time $St = 10$. For unstratified shear flow
with $Ri = 0$, the Lagrangian and Eulerian accelerations are almost decorrelated as indicated by $r= 0.0284$. With increasing
stratification strength, the Pearson product-moment correlation coefficient increases monotonically.
A high value of $r = 0.6634$ is observed for $Ri = 1$. This can be explained by the decreasing importance of the nonlinearity, quantified by the decreasing Reynolds number.

\begin{figure*} % 06
\setlength{\unitlength}{0.45\textwidth}
\includegraphics[bb=50 50 302 251,width=1\unitlength,angle=0]{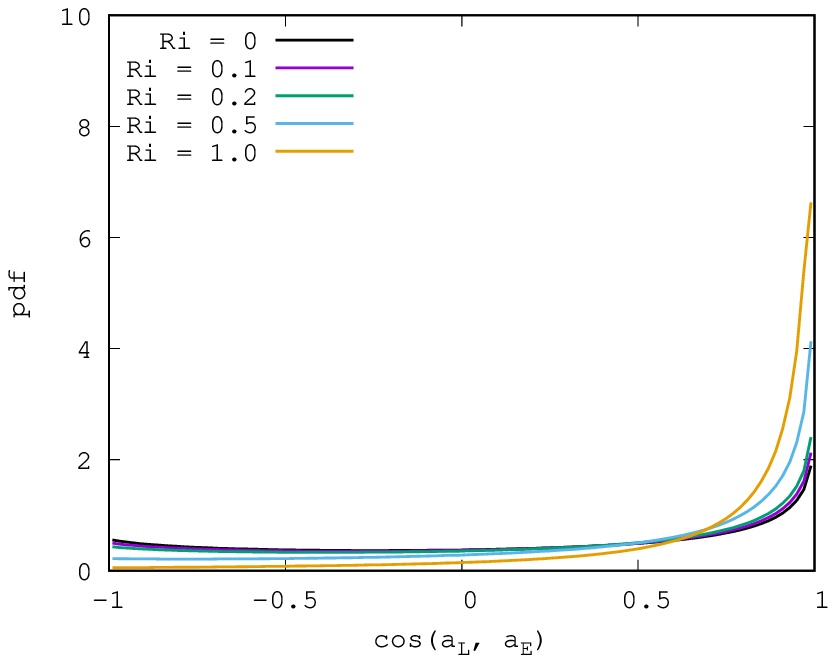}
\includegraphics[bb=50 50 302 251,width=1\unitlength,angle=0]{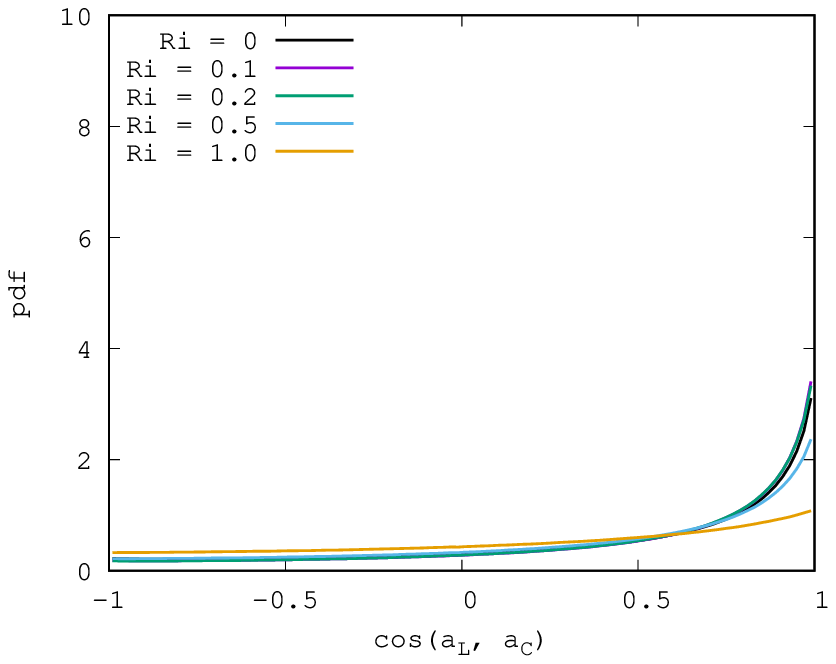}\\
\includegraphics[bb=50 50 302 251,width=1\unitlength,angle=0]{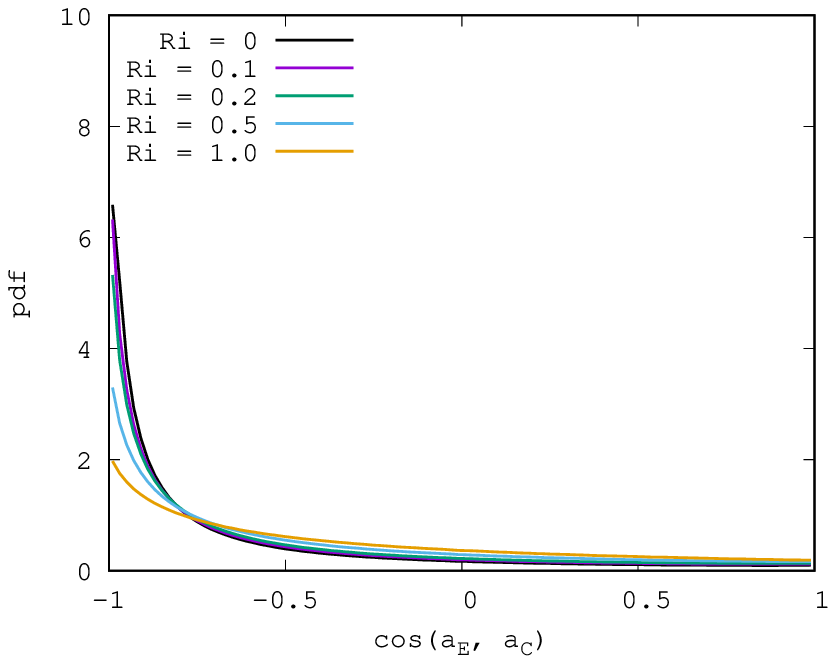}
\includegraphics[bb=50 50 302 251,width=1\unitlength,angle=0]{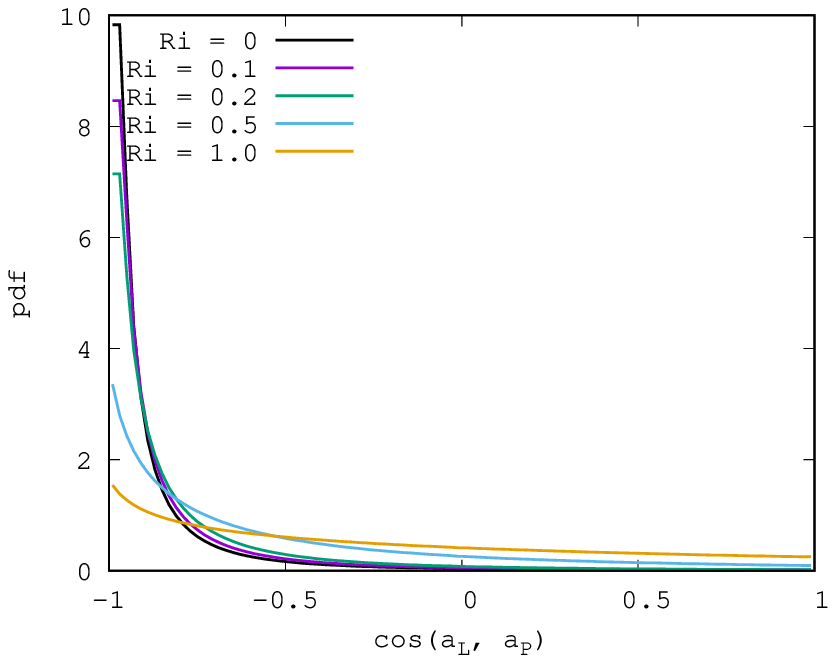}\\
\vspace{1.0cm}
\caption{Pdfs of the cosine of the angles between different acceleration contributions: 
         $\cos (a_L, a_E)$ (top, left), $\cos (a_L, a_C)$ (top, right), $\cos (a_E, a_C)$ (bottom, left),
         and $\cos (a_L, a_P)$ (bottom, right) for different $Ri$ at nondimensional time $St=10$.}
\label{fig_angstat}
\end{figure*}

To provide further information, geometrical statistics are performed and the vector alignment properties of the different acceleration contributions are studied, as introduced in section~\ref{sec:geostat}. 
For different Richardson numbers we consider the pdf of the cosine of the angle of two acceleration vectors, shown in Figure~\ref{fig_angstat}, the mean value of the cosine as quantitative measure and the correlation coefficient, assembled in table~\ref{tab_geostat}. The choice of the cosine is motivated by the fact that for random fields in 3D the cosine of the angle is uniformly distributed and not the angle itself.
Figure~\ref{fig_angstat} (bottom, left) shows that a strong anti-alignment of the Eulerian acceleration ${\bm a}_E$ and the convective term ${\bm a}_C = {\bm N}$ reflected in a peak in the pdf at $\cos= -1$, corresponding to an angle of $180$ degrees.
This explains why the Lagrangian acceleration is smaller than the Eulerian one and also smaller than the convective term, as the anti-alignment implies that the two vectors ${\bm a}_E$ and  ${\bm a}_C = {\bm N}$ are anti-parallel.
The Lagrangian acceleration ${\bm a}_L$  and the pressure gradient $\bm \Pi$ in Fig.~\ref{fig_angstat} (bottom, right) even show a stronger anti-alignment, which confirms that the negative pressure gradient is the driving force of the flow dynamics.
In both cases the anti-alignment is most pronounced for $Ri=0$ and becomes weaker for increasing Richardson numbers.
This can be further quantified by mean values of the cosine of the angle and also the correlation coefficient between the two vectors, the results are given in table~\ref{tab_geostat}.
%{\color{red}\it I would suggest to discuss one or two values in the table and to compare with the values of \citet{tsinvy2001}.}

The Lagrangian acceleration is positively aligned with the Eulerian one (Fig.~\ref{fig_angstat}, top, left) and also with the convective acceleration (Fig.~\ref{fig_angstat}, top, right). For ${\bm a}_E$ this alignment becomes stronger with increasing Richardson number, while for ${\bm a}_C$ this becomes weaker, as the nonlinear term diminishes.
These results are consistent with those in \citet{tsinvy2001} obtained for isotropic turbulence, in the case of $Re_{\lambda} = 141$ (as compared to our value for unstratified shear flow $157$). %(or 243). 
For example for the average cosine of the angle between $a_L$ and $a_E$ \citet{tsinvy2001}  reports a value $0.105$ (for unstratified shear flow we find $0.162$), for $a_L$ and $a_C$ 0.353 (0.420) and $a_E$ and $a_C$ -0.762 (-0.657).
%\citet{tsinvy2001} also mentions that this behavior was observed in experiments and geophysical flows. We could cite some references.

%{\it What about Taylor hypothesis? The alignment goes into the right direction and would support the Taylor hypothesis (${\bm a}_L << {\bm a}_E \approx {\bm a}_C$), but according to \citet{lin1953} it does not hold for shear flows. The problem is that we have indeed ${\bm a}_L < {\bm a}_E$ but it is the same order of magnitude, the ratio is 0.74 to 0.9 and Tsinober found smaller values, 0.33 for $256^3$, but normally $<<$ should be one order of magnitude. To be discussed. Maybe also a Reynolds number effect.}

Let us also mention that \citet{tsinvy2001} showed that Gaussian random fields satisfy similar alignment properties for ${\bm a}_E$ and ${\bm a}_C = {\bm N}$ and concluded that this is essentially a kinematic effect.

% figure 8

\begin{figure*} % 07
\setlength{\unitlength}{0.45\textwidth}
\includegraphics[bb=50 50 302 251,width=1\unitlength,angle=0]{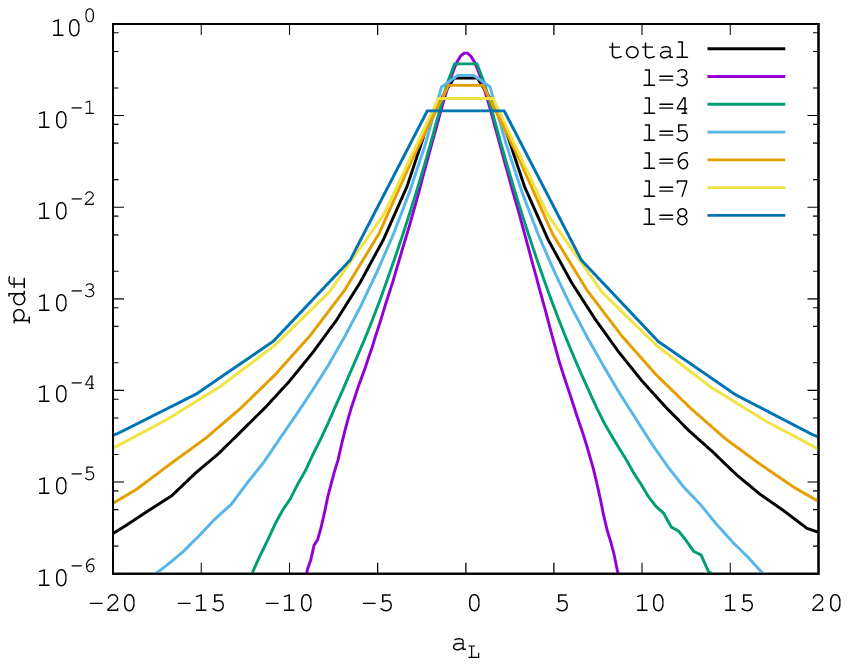}
\includegraphics[bb=50 50 302 251,width=1\unitlength,angle=0]{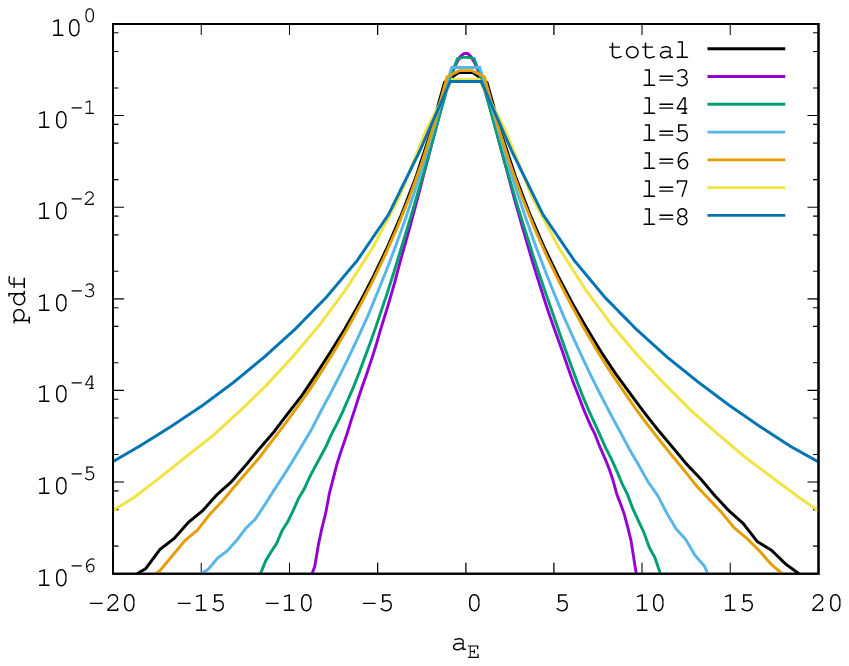}\\
\includegraphics[bb=50 50 302 251,width=1\unitlength,angle=0]{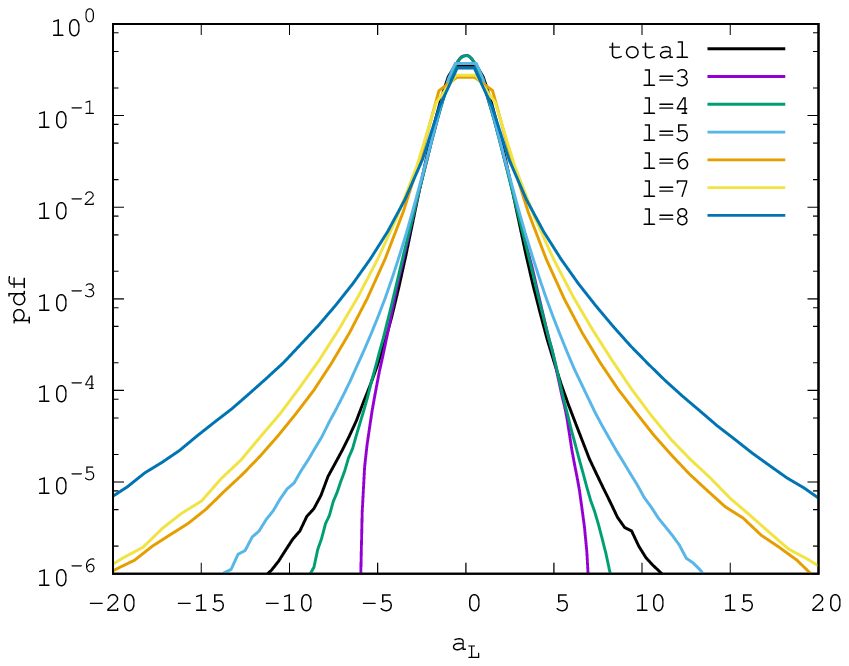}
\includegraphics[bb=50 50 302 251,width=1\unitlength,angle=0]{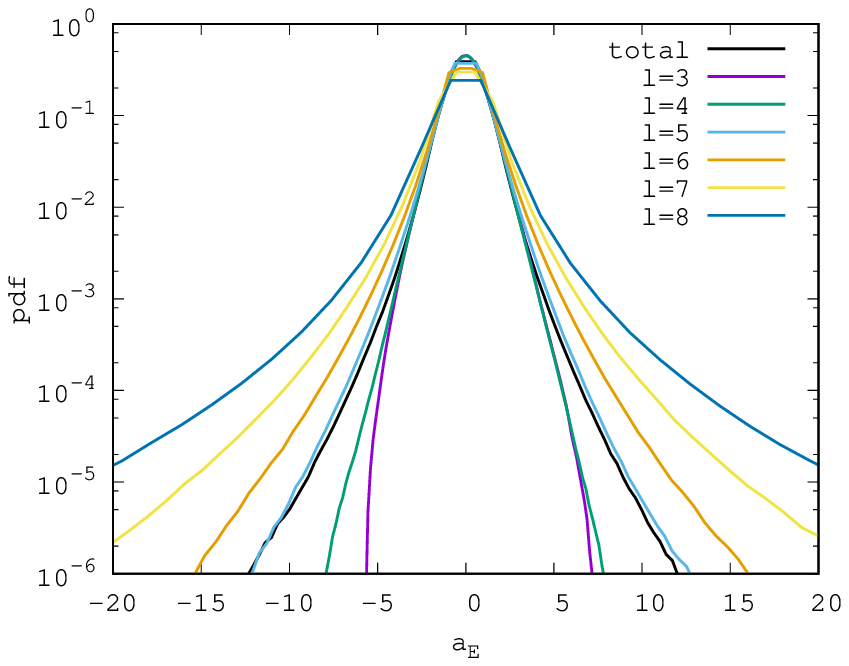}\\
\vspace{1.0cm}
\caption{Scale-dependent normalized pdfs of Lagrangian acceleration $\bm a_L$ (left) and Eulerian
         acceleration $\bm a_E$ (right) for Richardson numbers $Ri=0.1$ (top) and $Ri=1$ (bottom) at
         nondimensional time $St=10$. Note that the flat sections in the pdfs around zero are artifacts of an even number of bins chosen in the computation of the pdfs. Additionally, pdfs of quantities with large variances only show fewer bins in the figure.}
\label{fig_svpdf}
\end{figure*}

Figure \ref{fig_svpdf} presents the scale-dependent pdfs of the Lagrangian acceleration $\bm a_L$
(left) and Eulerian acceleration $\bm a_E$ (right) for two cases with $Ri=0.1$ (top) and $Ri=1$
(bottom) at nondimensional time $St=10$. The pdfs have stretched-exponential shapes and the tails
become heavier with increasing scale index $j$ or decreasing scale of the turbulent motion. For the
weakly stratified case with $Ri=0.1$, the tails of the Lagrangian acceleration are generally heavier
as those of the Eulerian acceleration. For the strongly stratified case with $Ri=1$, however, the
tails of the Eulerian acceleration are generally heavier than those of their Lagrangian counterpart.
This observation reflects the trend for the total acceleration pdfs with increasing Richardson number
discussed above.

% tables 5 and 6

In order to quantify the above observations, scale-dependent statistics are provided in tables
\ref{tab_laea_0010} and \ref{tab_laea_0100} for two cases with Richardson numbers $Ri=0.1$ and $Ri=1$,
respectively. While the magnitude of the total Eulerian acceleration $a_E$ is larger than the
magnitude of the total Lagrangian acceleration $a_L$, the ordering is reversed for the accelerations
at some scales of the turbulent motion. For the case with $Ri=0.1$, the original ordering holds at
the scale with the largest magnitude, which is $j=6$ for both accelerations. At that scale, the pdfs
of the total accelerations are also most similar to the pdfs of the accelerations at that scale. For
the case with $Ri=1$, the ordering observed for the magnitudes of the total accelerations holds more
generally at different scales of the turbulent motion. The pdfs of the total accelerations are again
most similar at the scales with the largest magnitudes, which are $j=4$ for the Lagrangian acceleration
and $j=5$ for the Eulerian acceleration. The flatness of the accelerations generally increases with
scale index $j$, indicating more intermittency at the smallest scales of motion. Note that for the
Lagrangian acceleration, flatness values close to three are observed for the larger scales with $j=1$,
indicating that the Lagrangian acceleration at large scale is mainly determined by linear effects.

% figure 9

\begin{figure*} % 08
\setlength{\unitlength}{0.45\textwidth}
\includegraphics[bb=50 50 302 251,width=1\unitlength,angle=0]{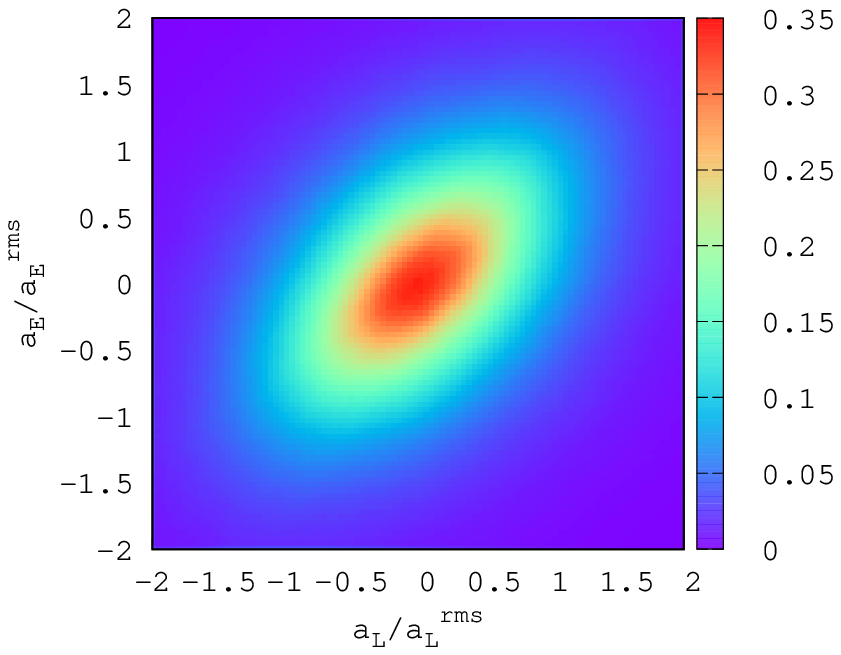}
\includegraphics[bb=50 50 302 251,width=1\unitlength,angle=0]{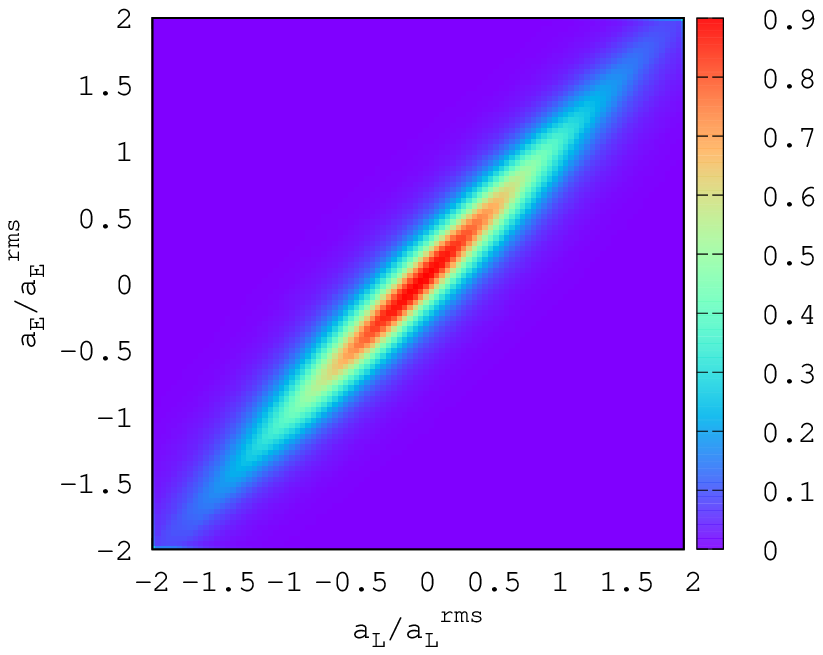}\\
\includegraphics[bb=50 50 302 251,width=1\unitlength,angle=0]{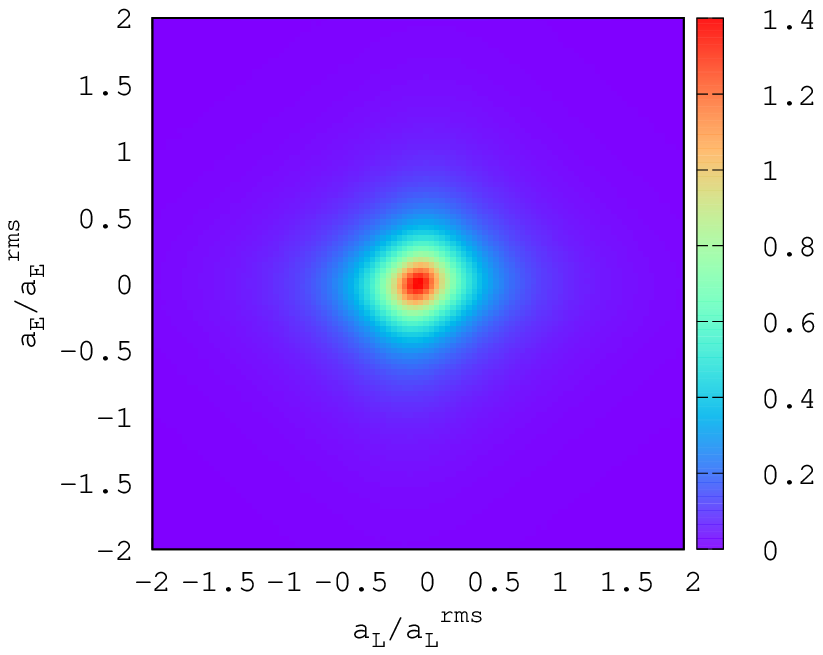}
\includegraphics[bb=50 50 302 251,width=1\unitlength,angle=0]{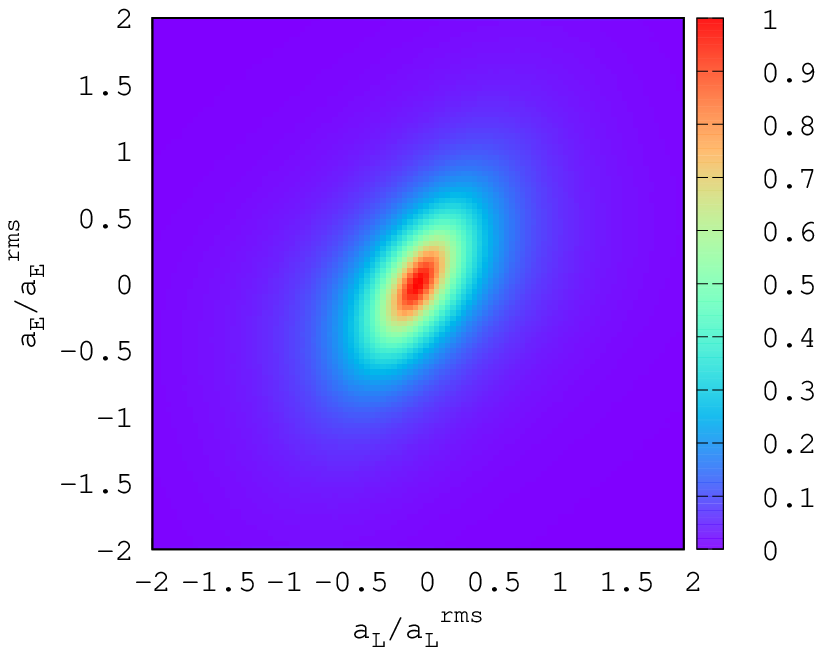}\\
\vspace{1.0cm}
\caption{Scale-dependent joint pdfs of Lagrangian acceleration $\bm a_L$ and Eulerian acceleration 
         $\bm a_E$ for Richardson numbers $Ri=0.1$ (left) and $Ri=1$ (right) and at large scale with 
         scale index $j=3$ (top) and at small scale with $j=7$ (bottom) at nondimensional time $St=10$ using a linear color scale.}
\label{fig_sjpdf}
\end{figure*}

Figure \ref{fig_sjpdf} shows the scale-dependent joint pdfs of the Lagrangian and Eulerian
accelerations for two cases with $Ri = 0.1$ (left) and $Ri = 1$ (right) as well as at large scale
with scale index $j=3$ (top) and at small scale with $j=7$ (bottom) at nondimensional time $St=10$.
Consistent with the observation for the total accelerations discussed above, the correlation
increases with stratification strength at the two scales shown. In addition, the correlation
decreases with increasing scale index $j$ or decreasing scale of the turbulent motion considered.

% table 4

This observation is shown more quantitatively using the Pearson product-moment correlation
coefficient in table \ref{tab_pearson}. The correlation coefficient tends to increase with
increasing Richardson number $Ri$. Similarly, at all Richardson numbers, the correlation
coefficient decreases with decreasing scale or increasing scale index $j$. 
%The only exception to this observation is obtained for strong stratification ($Ri \le 5$ ({\color{red} To be checked.})) and at small scale ($j \le 7$). 
The components at the largest scale index $j$ or smallest scale of motion are
characterized by very high flatness values. This indicates strong intermittency present in the
motion with the localized activity impacting the correlation coefficient.
Note that with increasing $Ri$, the Taylor micro-scale Reynolds number $Re_\lambda$ decreases. Starting from the same initial conditions, an increase of the Richardson number $Ri$ necessarily results in the decrease of the Taylor micro-scale Reynolds number $Re_\lambda$ due to the effect of stratification. Hence it is difficult to determine if the origin of the increased intermittency is due to the increased stratification or decreased $Re_\lambda$ as the two effects are linked.

\subsection{Lagrangian and Eulerian Time-Rates of Change of Fluctuating Density}

The time-rates of change of fluctuating density can also be defined using Lagrangian and Eulerian approaches as
\begin{equation}
s_L = \frac{\partial \rho}{\partial t} + \bm u \cdot \nabla \rho
\quad
\textrm{and}
\quad
s_E = \frac{\partial \rho}{\partial t},
\end{equation}
respectively.

% figures 13 and 14

\begin{figure*} % 09
\setlength{\unitlength}{0.45\textwidth}
\includegraphics[bb=50 50 302 251,width=1\unitlength,angle=0]{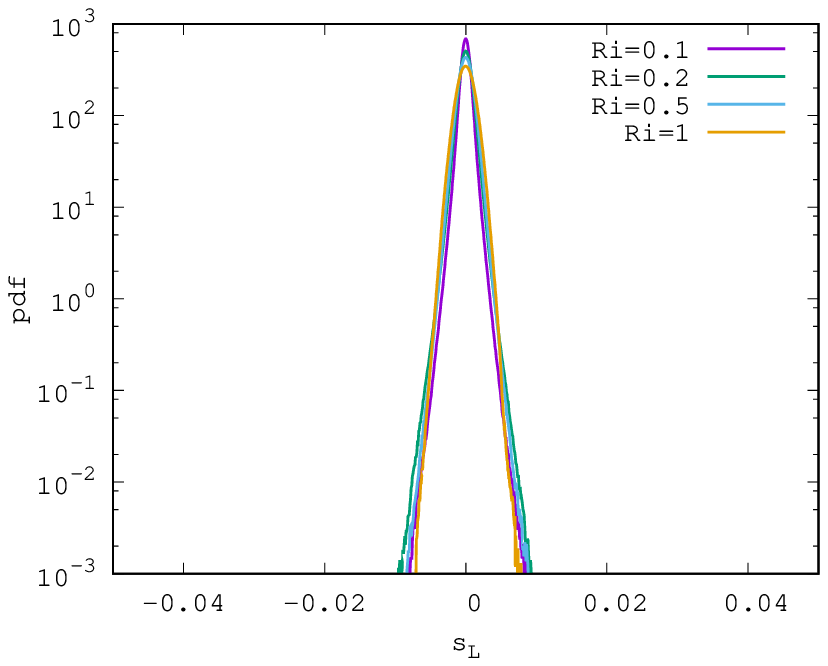}
\includegraphics[bb=50 50 302 251,width=1\unitlength,angle=0]{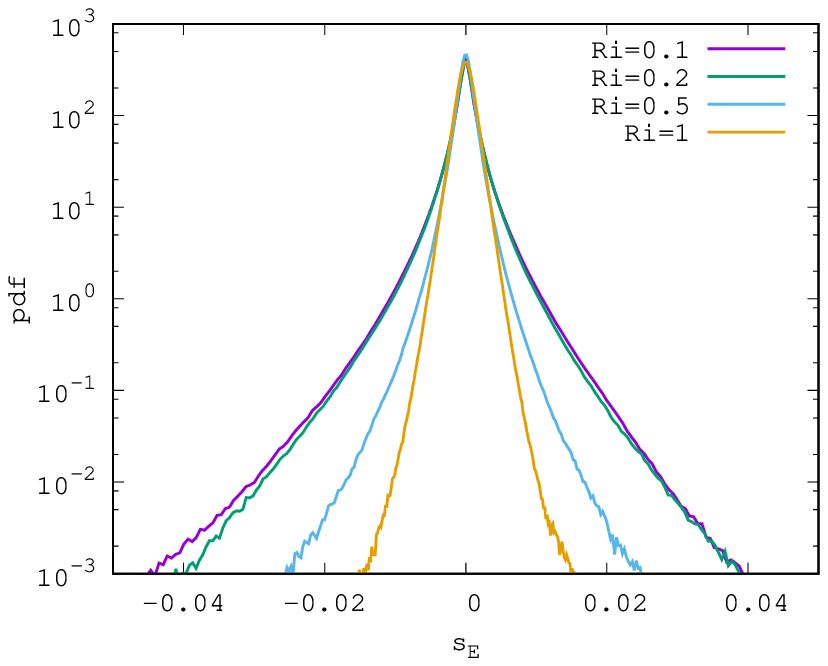}\\
\vspace{1.0cm}
\includegraphics[bb=50 50 302 251,width=1\unitlength,angle=0]{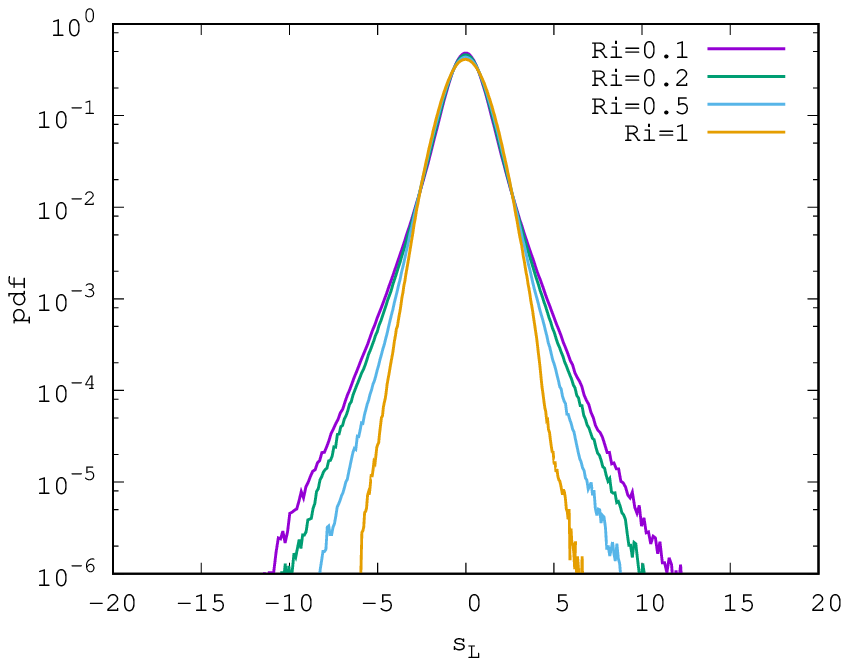}
\includegraphics[bb=50 50 302 251,width=1\unitlength,angle=0]{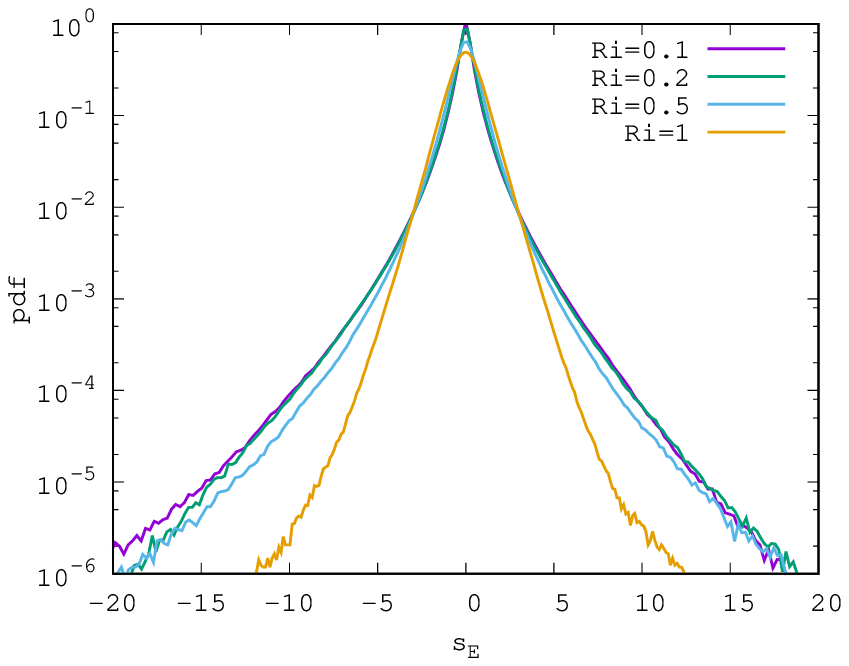}
\vspace{1.0cm}
\caption{Pdfs (top) and normalized pdfs (bottom) of Lagrangian time-rate of change of fluctuating 
         density $s_L$ (left) and Eulerian time-rate of change $s_E$ (right) at nondimensional time $St=10$.}
\label{fig_rpdf+fig_rpdfn}
\end{figure*}

Figure \ref{fig_rpdf+fig_rpdfn} (top) shows the pdfs of the Lagrangian time-rate of change of fluctuating density (left) and of
the corresponding Eulerian time-rate of change (right). The difference in the pdfs of the time-rates of change
is well more pronounced than the difference obtained for the accelerations. Figure \ref{fig_rpdf+fig_rpdfn} (bottom) shows the
normalized pdfs of the two time-rates of change. While the shape of the Eulerian time-rate of change pdf is
again found to be stretched-exponential, the Lagrangian time-rate of change pdf has a more Gaussian shape.
The extreme values of the Eulerian time-rate of change of fluctuating density are substantially larger than
those of the Lagrangian time-rate of change.

Tables \ref{tab_la} and \ref{tab_ea}, respectively, provide the dependence of the magnitudes of the Lagrangian
and Eulerian time-rates of change on the Richardson number $Ri$. Note that for $Ri=0$, the density is a passive
scalar (zero gravity) with a mean gradient. Again, the magnitude of $s_E$ always remains larger than the magnitude
of $s_L$, consistent with the findings for the accelerations.

The flatness of the Lagrangian and Eulerian time-rates of change are also given in tables \ref{tab_la} and \ref{tab_ea},
respectively. The flatness of $s_E$ is always larger than that of $s_L$ and their values generally decrease with
increasing $Ri$. 
For strong stratification, the flatness of the Lagrangian time-rate of change assumes values around three for $Ri=1$, while the Eulerian time-rate of change yields 
a value of $5.475$.
%three for the most strongly stratified case with $Ri=10$ ({\color{red} To be removed}).

\begin{figure*} % 10
\setlength{\unitlength}{0.45\textwidth}
\includegraphics[bb=50 50 302 251,width=1\unitlength,angle=0]{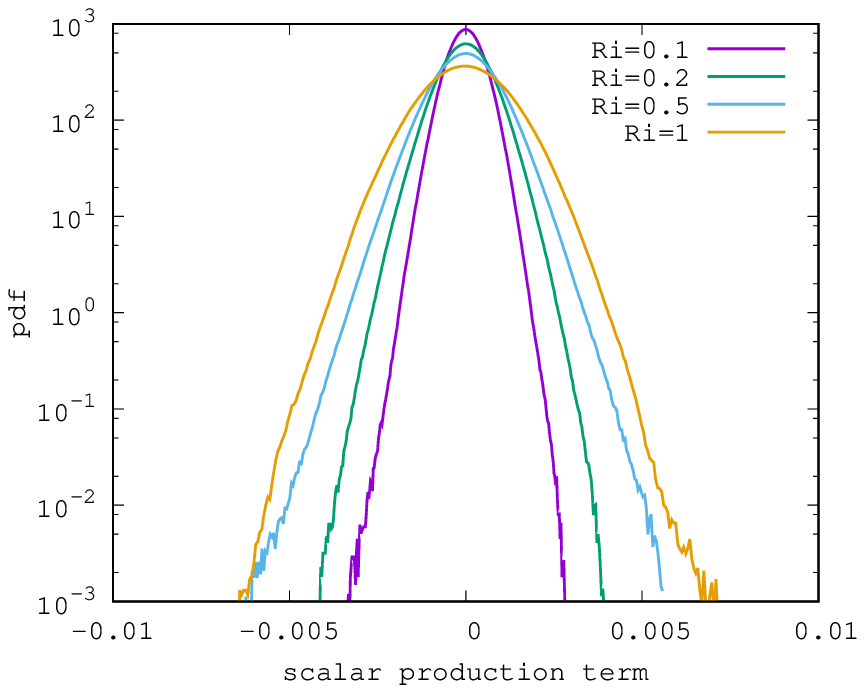}
\includegraphics[bb=50 50 302 251,width=1\unitlength,angle=0]{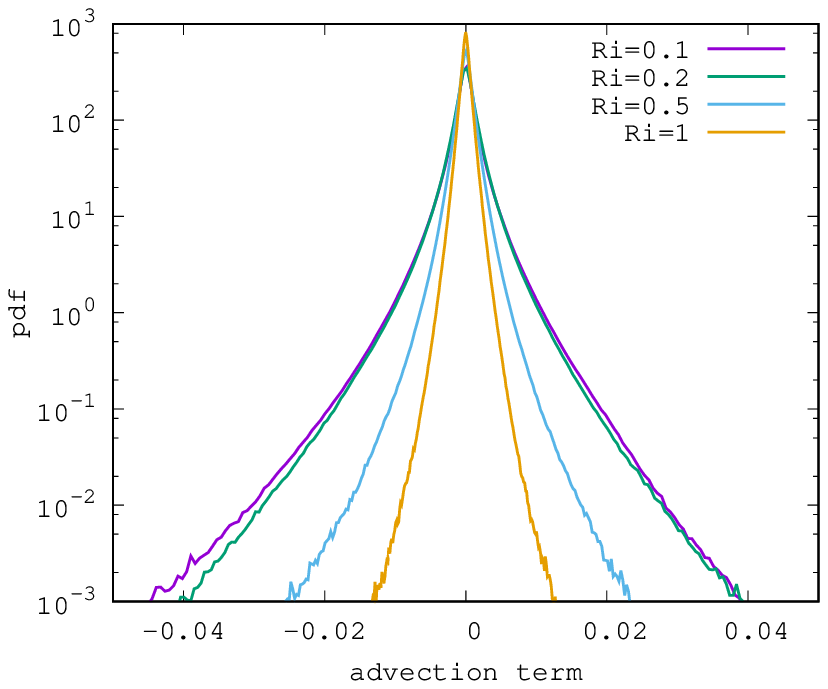}
\vspace{1.0cm}
\caption{Pdfs of the buoyancy (left) and advection (right) terms in the advection-diffusion equation for
         fluctuating density at nondimensional time $St=10$.}
\label{fig_addi}
\end{figure*}

Figure \ref{fig_addi} shows pdfs of the buoyancy term (left) and advection term (right) in the advection-diffusion
equation for fluctuating density. The buoyancy term pdf has a Gaussian shape as it is linearly related to the fluctuating
density. Its variance increases with increasing $Ri$, because the stratification rate $S_\rho$ increases. The more
Gaussian shape of the Lagrangian time-rate of change of fluctuating density can be explained by the lack of a quadratic
term in the advection-diffusion equation for fluctuating density. The large difference observed between the Lagrangian
and Eulerian time-rates of change of fluctuating density is due to the advection term.

%fertig bis hier.
\section{Conclusions}
\label{sec:conclusions}

A series of direct numerical simulations was performed in order to study the Lagrangian and Eulerian acceleration
properties in stably stratified turbulent shear flows. With increasing Richardson number $Ri$, the evolution of the
turbulent kinetic energy $K$ changes from growth to decay and the variances of the Lagrangian acceleration $\bm a_L$
and the Eulerian acceleration $\bm a_E$ decrease. The acceleration pdfs were observed to have a stretched-exponential
symmetric shape and the flatness decreases with increasing $Ri$. 

We studied the cancellation of Eulerian and convective accelerations of fluid particle using geometrical statistics of the vector quantities. We found a strong preference for the anti-alignment of both vectors, which decreases with the Richardson number.
This cancellation explains why the variance of the Lagrangian acceleration is smaller than its Eulerian counterpart and it supports, according to \citet{tsinvy2001}, who performed similar analyses for isotropic turbulence, the random Taylor hypothesis for shear flow which becomes however weaker with increasing stratification. 
Nevertheless, we do not find an order of magnitude difference in the acceleration variances, as predicted by \citet{tennekes1975} for isotropic turbulence and necessary so that the random Taylor hypothesis strictly holds. 
These findings are in agreement with \citet{lin1953}, who showed that Taylor's hypothesis does in general not hold for shear flow.
Analyzing the alignment properties of the scale-dependent contributions of the acceleration is an interesting perspective for future work, already in the context of isotropic turbulence. This would allow to check if the hypothesis holds for shear flows at least as small scales. 

An estimation of the variances of the Lagrangian and Eulerian accelerations has been derived from the Navier--Stokes equations which requires the ratio of potential to kinetic energy. A comparison of the estimation with results from the direct numerical simulations showed good agreement for the considered range of Richardson numbers.

The pdfs of the pressure-gradient and advection terms in the Navier--Stokes equation, which are both quadratic terms,
also have stretched-exponential shapes. The Lagrangian and Eulerian accelerations are mainly determined by the
pressure-gradient and advection terms, respectively. While the quadratic terms are dominant for small $Ri$, their
dominance is somewhat diminished for large $Ri$. The pdfs of the shear and buoyancy terms in the Navier--Stokes equation,
which are both linear terms, were observed to have a Gaussian shape. While the variance of the shear term decreases with
$Ri$, the variance of the buoyancy term increases with $Ri$.

In addition, the Lagrangian and Eulerian time-rates of change of fluctuating density and of fluctuating vorticity (see appendix~\ref{ap_vort}) were
considered. For both quantities, the Eulerian time-rates of change showed substantially larger extreme values than their
Lagrangian counterparts. Due to a lack of a quadratic term on the right-hand-side of the advection-diffusion equation for
fluctuating density, the pdf of the Lagrangian time-rate of change has an almost Gaussian shape, while the pdf of the
Eulerian time-rate of change was observed to have exponential to stretched-exponential shapes. For fluctuating vorticity
we found that the Lagrangian time-rate of change is mainly determined by the vortex streching and tilting term.

A scale-dependent analysis using orthogonal wavelet decomposition provided insight into the intermittency of the
Lagrangian and Eulerian accelerations. At small scales of the turbulent motion, the pdfs exhibit heavy tails, resulting
in very large flatness values and corresponding intermittency. The correlation between the Lagrangian and Eulerian
accelerations has likewise been analyzed and we found stronger correlation at large scales of turbulent motion as well
as with increasing Richardson number. At small scales this correlation is substantially reduced.

For rotating and sheared homogeneous turbulence, \citet{salhi2014} observed a dominance of linear terms in the cases
with strong rotation and the flatness of the Lagrangian acceleration assumes a value of about $3$. This observation
suggests that linear theory can accurately describe properties of such flows. In the present study, however, the
flatness never reaches values close to three, even for very large Richardson numbers. Hence, linear theory should not
yield agreement with direct numerical simulation results. Indeed, \citet{hanazaki2004} found important differences
between linear theory and the fully nonlinear evolution of homogeneous turbulence in stratified shear flows.

Perspectives for future work include a component-wise analysis of the Lagrangian and Eulerian acceleration, a more detailed scale-wise decomposition of the geometric properties of the accelerations, and corresponding terms for the vorticity evolution.

\begin{acknowledgments}

FGJ acknowledges the support from a University Professor Award from the University of San Diego and the hospitality
at Aix-Marseille Universit\'e. KS acknowledges financial support from Agence Nationale de la Recherche, project AIFIT (ANR-15-CE40-0019) and project CM2E (ANR-20-CE46-0010-01),
and the French Research Federation for Fusion Studies within the framework of the European Fusion Development Agreement (EFDA).

\end{acknowledgments}

%-------------------------appendix---------------------------------------------------------
\appendix

\section{Lagrangian and Eulerian Time-Rates of Change of Fluctuating Vorticity}
\label{ap_vort}
The Lagrangian and Eulerian time-rates of change of fluctuating vorticity $\bm \omega$ are defined as
\begin{equation}
\bm c_L=\frac{\partial \bm \omega}{\partial t} + \bm u \cdot \nabla \bm \omega
\quad
\textrm{and}
\quad
\bm c_E=\frac{\partial \bm \omega}{\partial t},
\end{equation}
respectively. This definition is analogous to the definition for the Lagrangian and Eulerian
accelerations in order to enable a comparison between the accelerations and vorticity time-rate
of change statistics. Again, the analysis is performed at the nondimensional time $St=10$.

% figures 10 and 11

\begin{figure*} % 11
\setlength{\unitlength}{0.45\textwidth}
\includegraphics[bb=50 50 302 251,width=1\unitlength,angle=0]{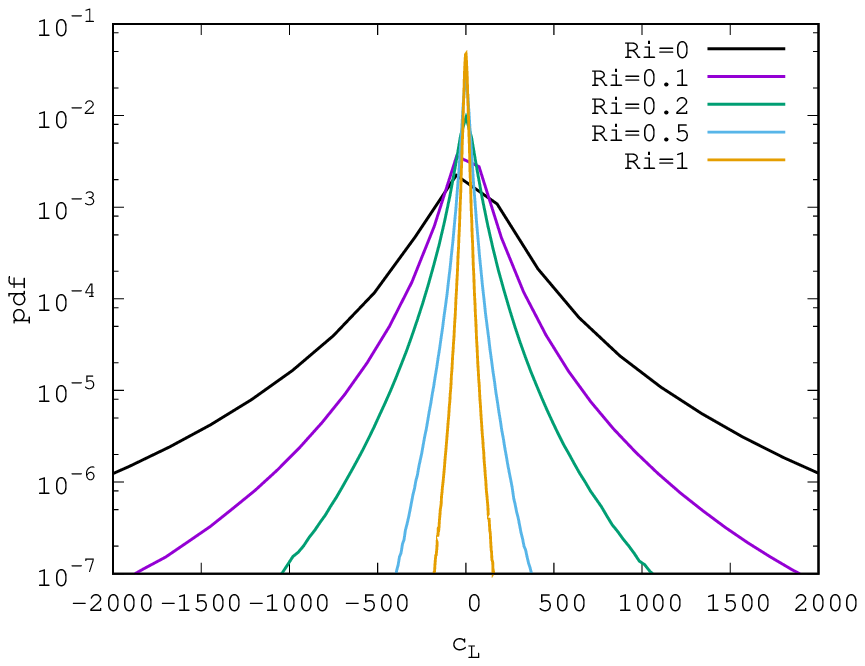}
\includegraphics[bb=50 50 302 251,width=1\unitlength,angle=0]{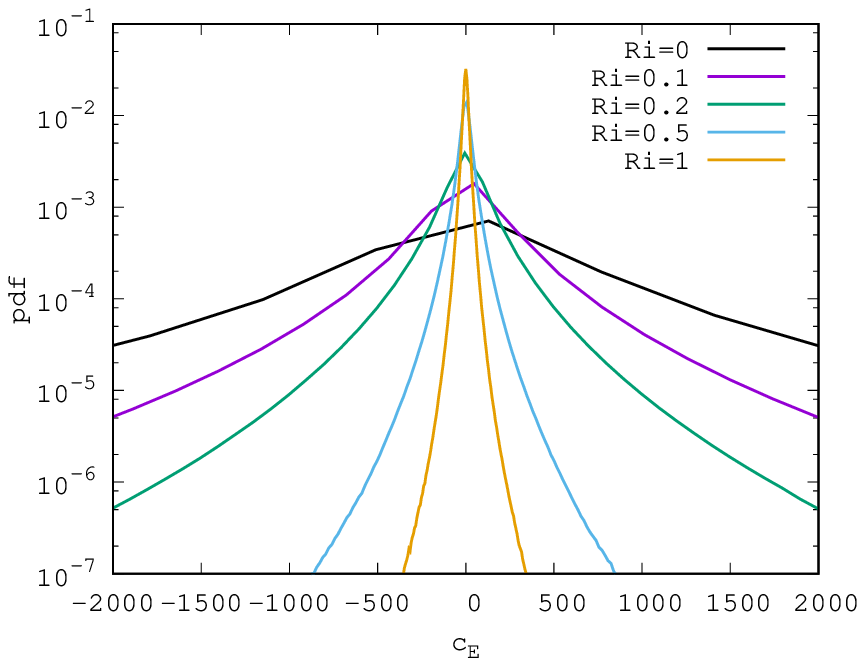}\\
\vspace{1.0cm}
\includegraphics[bb=50 50 302 251,width=1\unitlength,angle=0]{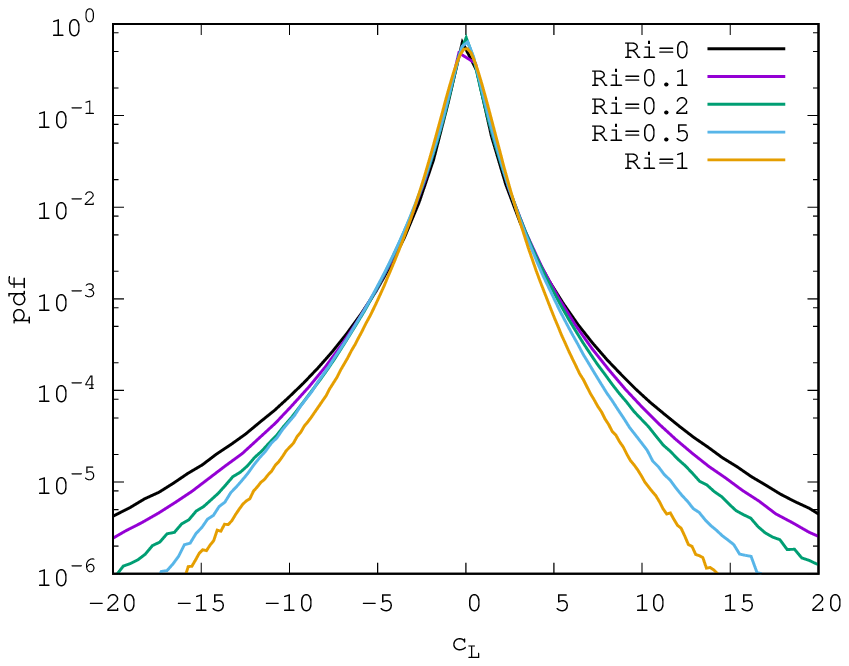}
\includegraphics[bb=50 50 302 251,width=1\unitlength,angle=0]{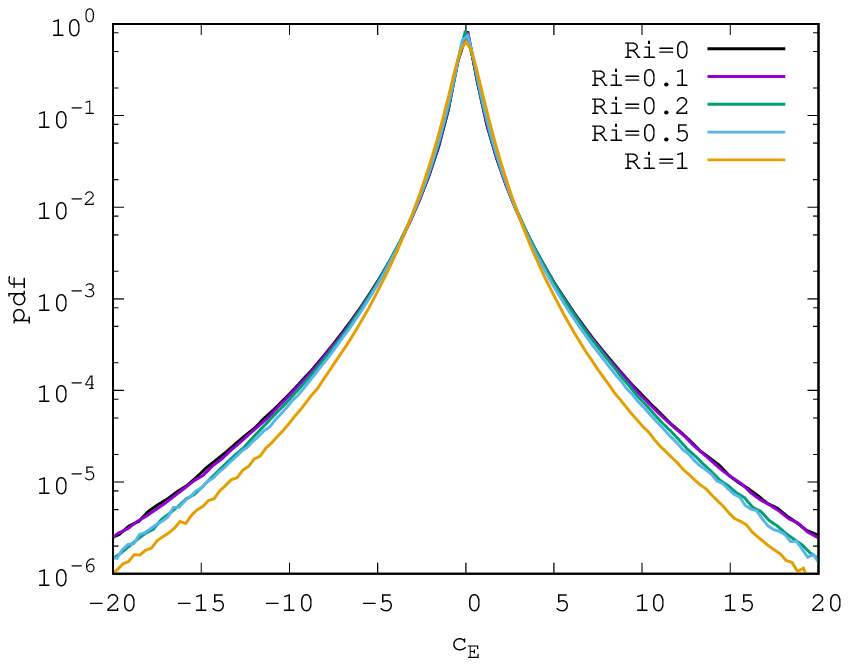}
\vspace{1.0cm}
\caption{Pdfs (top) and normalized pdfs (bottom) of Lagrangian time-rate of change of vorticity 
         $c_L$ (left) and Eulerian time-rate of change $c_E$ (right) at nondimensional time $St=10$. Note that pdfs for the vector quantities are shown.}
\label{fig_wpdf+fig_wpdfn}
\end{figure*}

Figure \ref{fig_wpdf+fig_wpdfn} (top) shows the probability distribution functions (pdfs) of the Lagrangian time-rate of
change $\bm c_L$ (left) and of the Eulerian time-rate of change $\bm c_E$ (right). Similar to the accelerations,
pdfs with stretched-exponential shapes are observed for both time-rates of change and a strong and similar
influence on the Richardson number $Ri$ is obtained. Again, stronger extreme values are obtained for the
Eulerian time-rate of change, but the difference to the Lagrangian time rate of change is much more pronounced
here as compared to the accelerations. Figure \ref{fig_wpdf+fig_wpdfn} (bottom) shows the normalized pdfs of the two time-rates of
change. Again, for a core region of about five standard deviations, both the Lagrangian and Eulerian time-rates
of change have an approximately similar shape. For small Richardson numbers $Ri$, the tails of the Lagrangian
time-rate of change are heavier than those of their Eulerian counterparts. However, this ordering is reversed at
larger $Ri$.

% tables 1 and 2 continued

The magnitudes of the Lagrangian and Eulerian vorticity time-rates of change are given in tables \ref{tab_la}
and \ref{tab_ea}, respectively. Similar to the magnitudes of the accelerations, the magnitudes of both time-rates
of change decrease with increasing $Ri$ and the variance of $\bm c_E$ remains always larger than the variance of
$\bm c_L$. This difference in the magnitudes for the vorticity time-rate of change pdfs is much more pronounced
than that of the accelerations. The heavier tails observed for the pdf of $\bm c_L$ as compared to $\bm c_E$ at
small $Ri$ results in a larger flatness of the Lagrangian time-rate of change pdf as compared to its Eulerian
counterpart. Again, the ordering of the flatness values is reversed at larger $Ri$. While the flatness values
decrease with increasing $Ri$, the flatness is again observed to level off at a value of approximately 5, well
above the value of 3 expected for a Gaussian pdf. Hence, some nonlinearity is still present even in the case
of strongly suppressed turbulence in strongly stratified flows.

% figure 12

\begin{figure*} % 12
\setlength{\unitlength}{0.45\textwidth}
\includegraphics[bb=50 50 302 251,width=1\unitlength,angle=0]{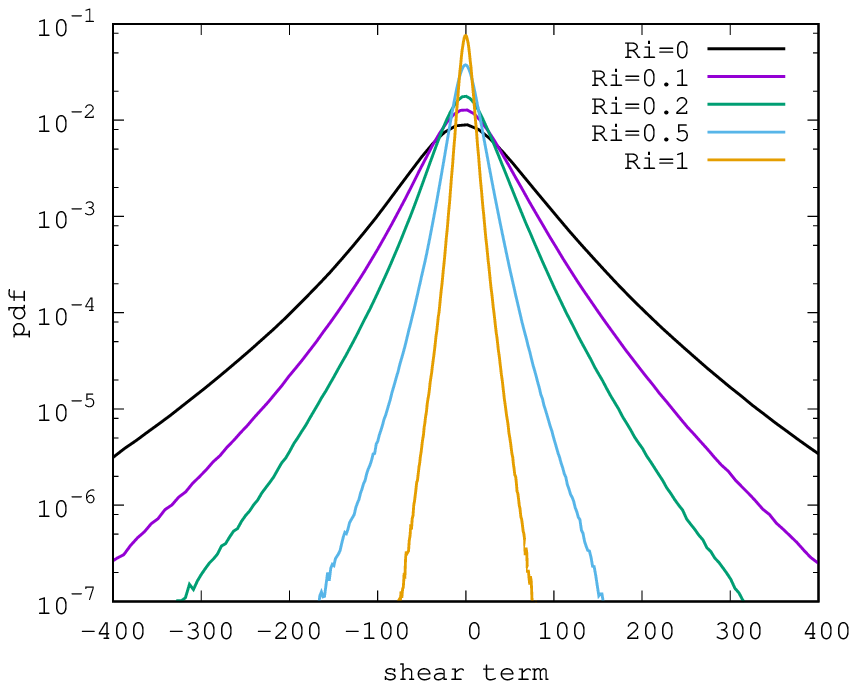}
\includegraphics[bb=50 50 302 251,width=1\unitlength,angle=0]{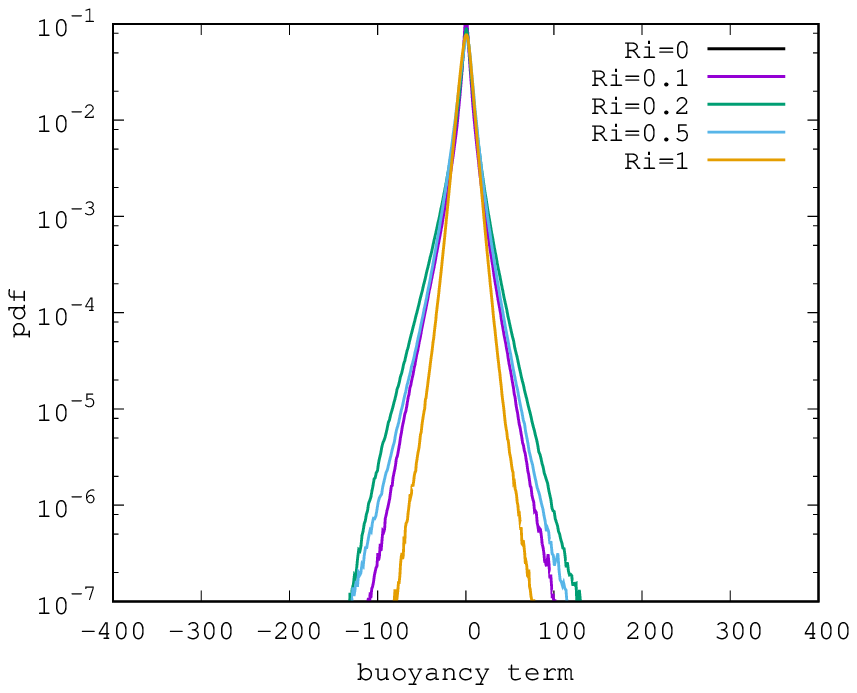}\\
\includegraphics[bb=50 50 302 251,width=1\unitlength,angle=0]{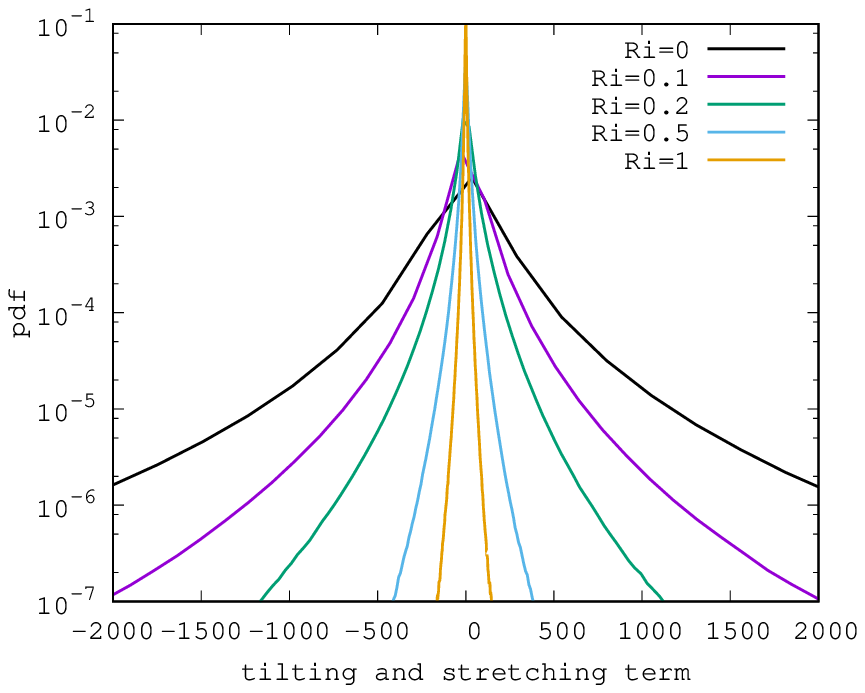}
\includegraphics[bb=50 50 302 251,width=1\unitlength,angle=0]{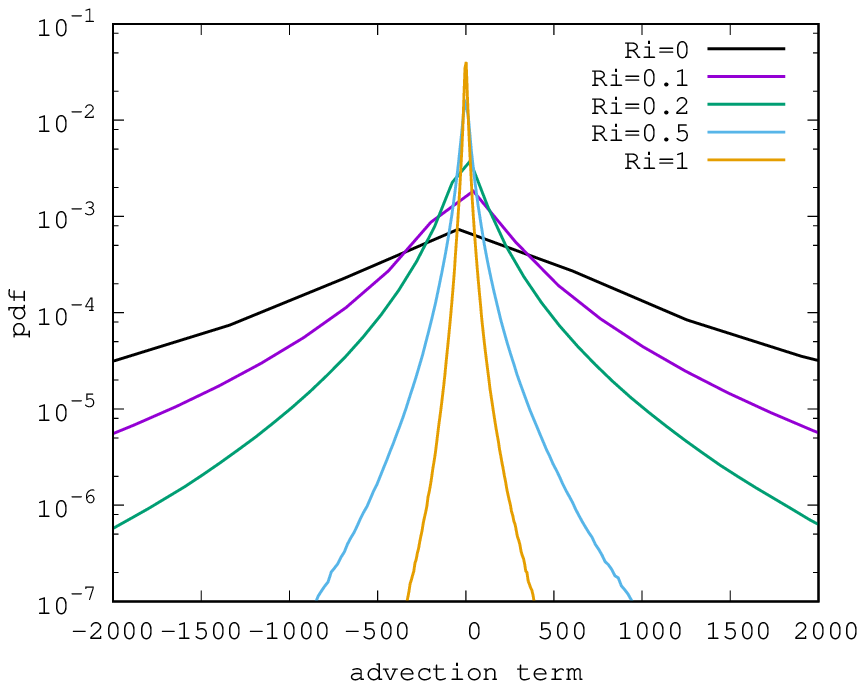}\\
\vspace{1.0cm}
\caption{Pdfs of the shear (top, left), buoyancy (top, right), vortex tilting and stretching 
         (bottom, left), and advection (bottom, right) terms in the vorticity equation at 
         nondimensional time $St=10$.}
\label{fig_vort}
\end{figure*}

Figure \ref{fig_vort} shows pdfs of the shear term (top, left), the buoyancy term (top, right), the vortex
tilting and stretching term (bottom, left), and the advection term (bottom, right) in the vorticity equation.
The shear and buoyancy terms depend linearly on the curl of fluctuating velocity components and fluctuating
density, respectively. Similarly to the respective terms in the Navier--Stokes equation, the magnitude of the
shear term decreases with increasing $Ri$ and the magnitude of the buoyancy term increases. The pdfs of the
vortex tilting and stretching term and the advection term show stretched-exponential shapes due to the quadratic
nature of the terms. The magnitudes of both terms decrease with increasing $Ri$. For small $Ri$, the vortex
tilting and stretching term as well as the advection term clearly dominate the shear and buoyancy terms, but
this dominance again is reduced for large $Ri$. In the case of vorticity, the vortex tilting and stretching
term is the generally dominant contribution to the Lagrangian time-rate of change, while the advection term
is important for the Eulerian time-rate of change.

%-------------------------appendix---------------------------------------------------------

\section{Lagrangian and Eulerian component pdfs}

% figure

\begin{figure*} % 13
\setlength{\unitlength}{0.45\textwidth}
\includegraphics[bb=50 50 302 251,width=1\unitlength,angle=0]{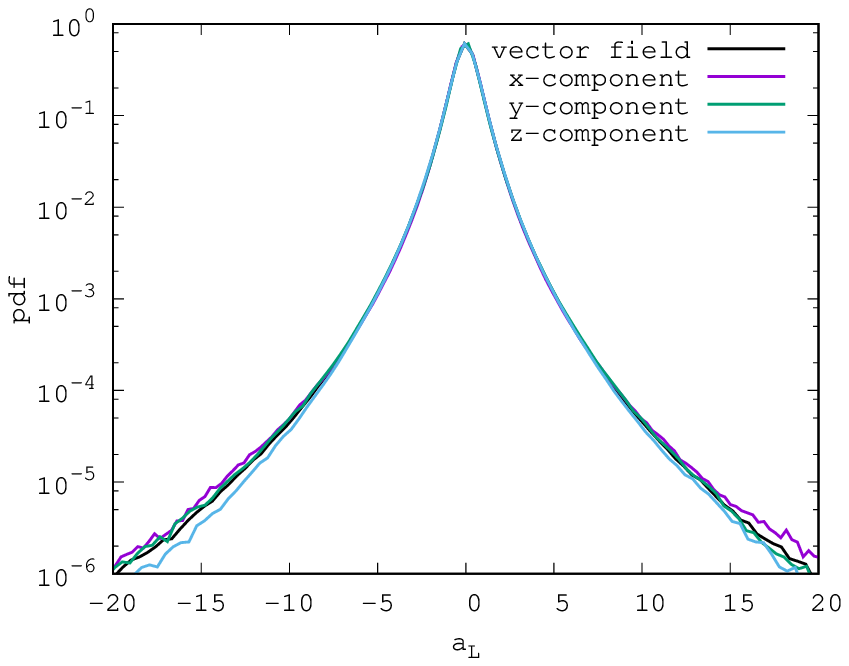}
\includegraphics[bb=50 50 302 251,width=1\unitlength,angle=0]{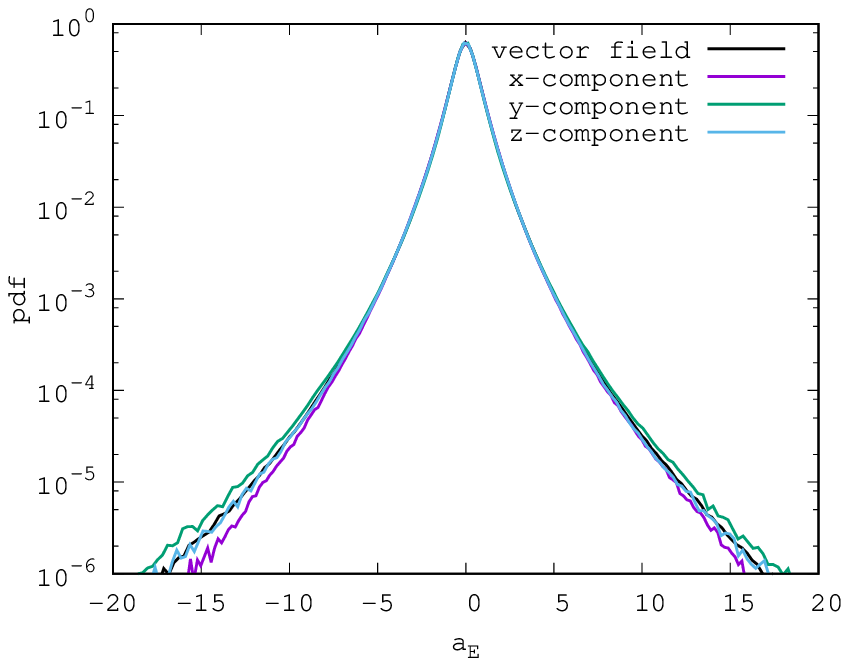}\\
\includegraphics[bb=50 50 302 251,width=1\unitlength,angle=0]{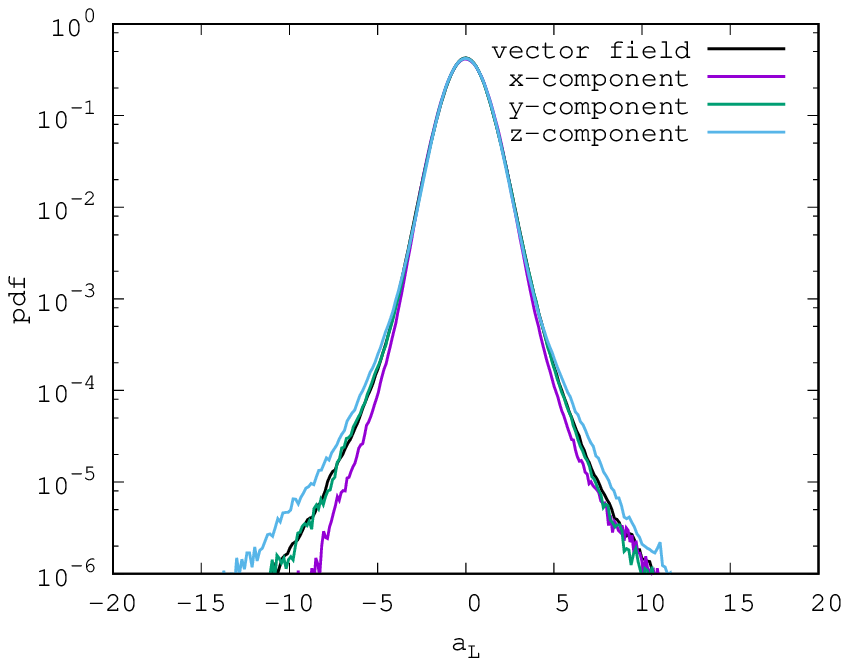}
\includegraphics[bb=50 50 302 251,width=1\unitlength,angle=0]{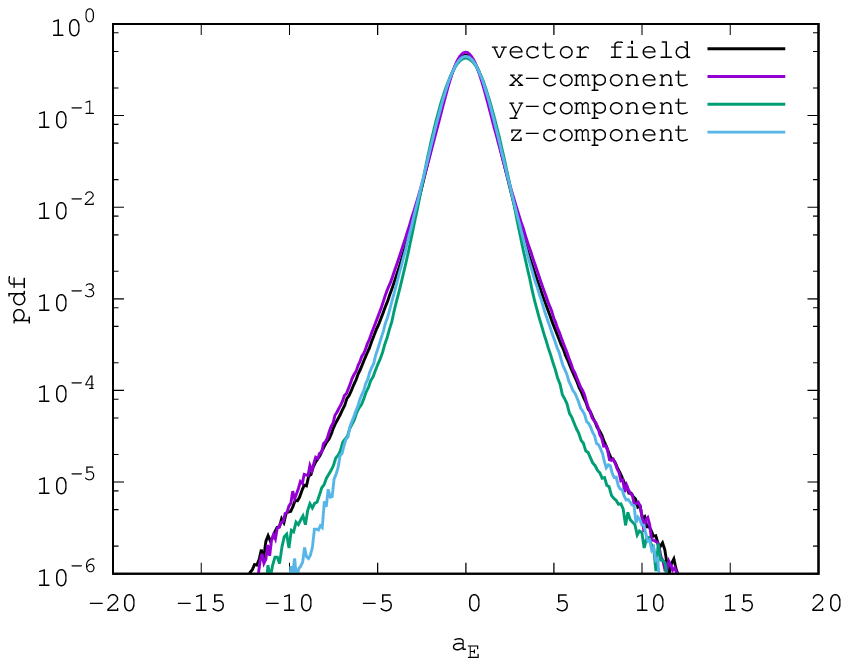}\\
\vspace{1.0cm}
\caption{Comparison of normalized vector pdfs with their corresponding normalized component pdfs for
         Lagrangian acceleration (left) and Eulerian acceleration (right) for $Ri=0.1$ (top) and 
         $Ri=1$ (bottom) at nondimensional time $St=10$.}
\label{fig_components}
\end{figure*}

While the main manuscript exclusively discusses the properties of vector pdfs, this appendix presents component pdfs of the Lagrangian and Eulerian accelerations in order to address their anisotropy in turbulent stratified shear flow.
Figure \ref{fig_components} compares the vector pdfs with their $x-$, $y-$, and $z-$component pdfs for the
Lagrangian acceleration (left) and Eulerian acceleration (right) for two cases with weak stratification with
$Ri = 0.1$ (top) and with strong stratification with $Ri = 1$ (bottom). All pdfs show similar shapes and
the flow anisotropy is reflected in the variances. 
A similar observation holds for the pdfs of Lagrangian and Eulerian time-rate of change of fluctuating vorticity (not shown here).

%
% tables 1 and 2 continued
%
The ratios of the component variances to the corresponding vector variances of the Lagrangian and Eulerian
accelerations are given in tables \ref{tab_la} and \ref{tab_ea}, respectively. For small Richardson numbers
$Ri$, the variance ratios show an almost equipartition between the three components for both the Lagrangian
and Eulerian accelerations. For large $Ri$, however, the ratio of the vertical variances to the vector variances
gains due to the direct impact of the buoyancy term in the vertical component of the Navier--Stokes equation. 
Figure 2 shows that the ratio of potential to kinetic energy increases with increasing $Ri$. Hence the buoyancy term impacts particularly the vertical component of the accelerations.

%--------------

%----------------------------------------------------------------------------------

\bibliography{PRFbib.bib}

%------------------------ figures ---------_----------------------------------------
\clearpage

\clearpage

\clearpage

\clearpage

\clearpage

\clearpage

\clearpage

\clearpage

\clearpage

\clearpage

\clearpage

\clearpage

\clearpage

%----------------------------- tables -----------------------------

\begin{table*}  % table 1
\caption{
Overview of the simulation cases, including the Richardson number $Ri$, the shear rate $S$, the Brunt--V\"ais\"al\"a frequency $N$, the Taylor-micro scale Reynolds number $Re_\lambda$, The viscosity $\nu$, the turbulent velocity fluctuation  $q$, the dissipation rate of kinetic energy $\epsilon$, the cut-off wavenumber $k_{max} \eta$, the overturning scale $L_{overturn}$, the Ellison scale $L_{Ellison}$, the Ozmidov scale $L_{Ozmidov}$, the Taylor microscale $\lambda$, and the Kolmogorov scale $\eta$. All values are given at nondimensional time $St=10$. All symbols are defined in the text. 
}
\label{tab_param}
\begin{ruledtabular}
\begin{tabular}{lrrrrrrrr}
$Ri$              &      0  &    0.1  &    0.2  &    0.5  &     1   \\ % &     2   &     5   &    10   \\
\hline
$S$               & 5.3345  & 5.3345  & 5.3345  & 5.3345  & 5.3345  \\ % & 5.3345  & 5.3345  & 5.3345  \\
$N$               & 0.0000  & 1.6869  & 2.3856  & 3.7720  & 5.3345  \\ % & 7.5441  & 11.9282 & 16.8691 \\
$Re_{\lambda}$    & 156.90  & 104.08  & 76.60   & 42.84   & 32.83   \\ % & 34.40   & 37.71   & 40.34   \\
$\nu$             & 0.0010  & 0.0010  & 0.0010  & 0.0010  & 0.0010  \\ % & 0.0010  & 0.0010  & 0.0010  \\
$q$               & 1.5856  & 1.0779  & 0.7969  & 0.4428  & 0.3028  \\ % & 0.2814  & 0.2975  & 0.3183  \\
$\epsilon$     & 1.2838  & 0.6230  & 0.3436  & 0.1048  & 0.0390  \\ % & 0.0265  & 0.0275  & 0.0315  \\
$k_{max} \eta$    & 1.1992  & 1.4368  & 1.6673  & 2.2437  & 2.8723  \\ % & 3.1643  & 3.1339  & 3.0289  \\
$L_{overturn}$    & 3.1052  & 2.0101  & 1.4728  & 0.8288  & 0.7119  \\ % & 0.8410  & 0.9563  & 1.0223  \\
$L_{Ellison}$     & 0.2619  & 0.1835  & 0.1324  & 0.0619  & 0.0312  \\ % & 0.0204  & 0.0132  & 0.0098  \\
$L_{Ozmidov}$     & N/A     & 0.3603  & 0.1591  & 0.0442  & 0.0160  \\ % & 0.0079  & 0.0040  & 0.0026  \\
$\lambda$         & 0.09895 & 0.09656 & 0.09613 & 0.09675 & 0.10841 \\ % & 0.12227 & 0.12677 & 0.12673 \\
$\eta$            & 0.00528 & 0.00633 & 0.00735 & 0.00988 & 0.01265 \\ % & 0.01394 & 0.01381 & 0.01334 \\
\end{tabular}
\end{ruledtabular}
\end{table*}

\begin{table*}  % table 2
\caption{Lagrangian acceleration statistics at nondimensional time $St=10$ showing the rms of the Lagrangian acceleration $a_L$, the flatness $Fl_{a_L}$ and the ratio of the component-wise variances and the total variance. The variance and flatness values of the time-rates of change of fluctuating density $s_L$ and $Fl_{s_L}$ and fluctuating vorticity $c_L$ and $Fl_{c_L}$ are likewise given.}
\label{tab_la}
%\begin{center}
\begin{ruledtabular}
\begin{tabular}{lrrrrrrrr}
$Ri$             &     0   &   0.1   &   0.2   &   0.5   &    1    \\ % &    2    &    5    &   10    \\
\hline
$a_L$            &  19.427 &   9.833 &   5.519 &   1.808 &   0.942 \\ % &   0.954 &   1.317 &   1.815 \\
$Fl_{a_L}$       &  27.814 &  26.041 &  13.115 &   9.364 &   4.111 \\ % &   3.480 &   3.737 &   3.938 \\
$a^2_{Lx}/a^2_L$ &   0.335 &   0.318 &   0.306 &   0.320 &   0.327 \\ % &   0.263 &   0.203 &   0.150 \\
$a^2_{Ly}/a^2_L$ &   0.325 &   0.327 &   0.331 &   0.356 &   0.390 \\ % &   0.512 &   0.569 &   0.607 \\
$a^2_{Lz}/a^2_L$ &   0.340 &   0.355 &   0.363 &   0.324 &   0.242 \\ % &   0.224 &   0.228 &   0.242 \\
$s_L$            & 0.00117 & 0.00070 & 0.00091 & 0.00101 & 0.00119 \\ % & 0.00200 & 0.00526 & 0.01096 \\
$Fl_{s_L}$       &   7.220 &   6.148 &   5.152 &   3.989 &   3.249 \\ % &   2.995 &   3.058 &   3.042 \\
$c_L$            & 490.484 & 232.508 & 128.472 &  44.395 &  19.725 \\ % &  13.761 &  15.074 &  19.171 \\
%kai: check notation of Fl: Fl [{\bm a}^j ]
$Fl_{c_L}$       &  65.574 &  53.766 &  21.168 &  14.123 &   9.177 \\ % &   5.155 &   4.732 &   5.247 \\
\end{tabular}
\end{ruledtabular}
\end{table*}

\begin{table*}  % table 3
\caption{Eulerian acceleration statistics at nondimensional time $St=10$ showing the rms of the Eulerian acceleration $a_E$, the flatness $Fl_{a_E}$ and the ratio of the component-wise variances and the total variance. The variance and flatness values of the time-rates of change fluctuating density $s_E$ and $Fl_{s_E}$ and of fluctuating vorticity $c_E$ and $Fl_{c_E}$ are likewise given.}
\label{tab_ea}
%\begin{center}
\begin{ruledtabular}
\begin{tabular}{lrrrrrrrr}
$Ri$             &     0     &   0.1    &   0.2.   &   0.5.   &    1     \\ % &    2     &    5     &   10     \\
\hline
$a_E$            &    26.100 &   11.689 &    6.258 &    1.975 &    1.075 \\ % &    1.047 &    1.404 &    1.898 \\
$Fl_{a_E}$       &    14.414 &   13.413 &   11.028 &    9.510 &    5.691 \\ % &    3.844 &    3.650 &    3.715 \\
$a^2_{Ex}/a^2_E$ &     0.302 &    0.320 &    0.339 &    0.408 &    0.434 \\ % &    0.322 &    0.247 &    0.186 \\
$a^2_{Ey}/a^2_E$ &     0.329 &    0.312 &    0.292 &    0.261 &    0.273 \\ % &    0.409 &    0.491 &    0.548 \\
$a^2_{Ez}/a^2_E$ &     0.369 &    0.368 &    0.369 &    0.331 &    0.293 \\ % &    0.269 &    0.262 &    0.265 \\
$s_E$            &   0.00577 &  0.00239 &  0.00231 &  0.00138 &  0.00129 \\ % &  0.00209 &  0.00537 &  0.01109 \\
$Fl_{s_E}$       &    24.992 &   24.624 &   21.960 &   16.053 &    5.475 \\ % &    3.742 &    3.438 &    3.264 \\
$c_E$            & 2,020.203 &  781.238 &  372.030 &   91.830 &   34.461 \\ % &   22.437 &   23.462 &   27.517 \\
$Fl_{c_E}$       &    37.813 &   35.619 &   27.855 &   24.893 &   17.724 \\ % &    8.437 &    7.180 &    6.642 \\
\end{tabular}
\end{ruledtabular}
\end{table*}

\begin{table*}  % table 4
\caption{Variance of the contributions to the linear term from the shear term $\Lambda_S^2$, the
         buoyancy term $\Lambda_B^2$, and viscous term $\Lambda_V^2$, an estimate for the variance
         of the linear term using the triangle inequality $\Lambda_{DNS}^2$, the variance of the velocity
         $q^2$, the ratio of potential to kinetic energies $K_\rho/K$, and an estimate for the linear
         term $\Lambda^2$ given in equation \ref{eqn_theory_valin} at nondimensional time $St=10$. }
\label{tab_est}
\begin{ruledtabular}
\begin{tabular}{lrrrrrrrr}
$Ri$              &      0   &    0.1   &    0.2  &    0.5  &     1   \\ % &     2   &     5   &    10   \\
\hline
$\Lambda_S^2$     &  14.5746 &   6.3796 &  3.2517 &  0.8430 &  0.3596 \\ % &  0.2764 &  0.3130 &  0.3404 \\
$\Lambda_B^2$     &   0.0000 &   0.2728 &  0.5681 &  0.7767 &  0.7886 \\ % &  1.3474 &  3.5150 &  7.7800 \\
$\Lambda_V^2$     &   6.8850 &   2.5158 &  1.1384 &  0.2568 &  0.0723 \\ % &  0.0355 &  0.0314 &  0.0348 \\
$\Lambda_{DNS}^2$ &  21.4596 &   9.1682 &  4.9581 &  1.8764 &  1.2205 \\ % &  1.6593 &  3.8594 &  8.1551 \\
$q^2$             &   2.5142 &   1.1619 &  0.6350 &  0.1961 &  0.0917 \\ % &  0.0792 &  0.0885 &  0.1013 \\
$K_\rho/K$        &   0.0000 &   0.0825 &  0.1571 &  0.2783 &  0.3021 \\ % &  0.2990 &  0.2791 &  0.2698 \\
$\Lambda^2$       &  23.8427 &  11.2910 &  6.5897 &  2.6358 &  1.6579 \\ % &  2.0975 &  4.3525 &  8.7371 \\
%$Re_\lambda$      & 156.90   & 104.08   & 76.60   & 42.84   & 32.83   & 34.40   & 37.71   & 40.34   \\
\end{tabular}
\end{ruledtabular}
\end{table*}

\begin{table*}  % table 5
\caption{Mean value of the cosine of the angle $\overline \cos$ and Pearson product-moment correlation coefficient $r$ between the Lagrangian accelerations ${\bm a}_L$, the Eulerian acceleration ${\bm a}_E$, the convective contribution ${\bm a}_C = {\bm N}$ and the pressure gradient ${\bm a}_C = \bm \Pi$ 
         at nondimensional time $St=10$. The correlation coefficient is determined for all three components
         of the vector fields. 
         }
\label{tab_geostat}
%\begin{center}
\begin{ruledtabular}
\begin{tabular}{lrrrrrrrr}
$Ri$                                    &     0   &    0.1  &    0.2  &    0.5  &     1   \\
\hline
$r ({\bm a}_L, {\bm a}_E)$              &  0.0284 &  0.0510 &  0.0882 &  0.2852 &  0.6634 \\
$r ({\bm a}_L, {\bm a}_C)$              &  0.5823 &  0.6205 &  0.6232 &  0.5493 &  0.2741 \\
$r ({\bm a}_E, {\bm a}_C)$              & -0.7961 & -0.7516 & -0.7241 & -0.6443 & -0.5378 \\
$r ({\bm a}_L, {\bm a}_P)$              & -0.9728 & -0.9545 & -0.9211 & -0.7014 & -0.2843 \\
\hline
$\overline \cos ({\bm a}_L, {\bm a}_E)$ &  0.1617 &  0.1957 &  0.2397 &  0.4347 &  0.7033 \\
$\overline \cos ({\bm a}_L, {\bm a}_C)$ &  0.4199 &  0.4579 &  0.4582 &  0.3749 &  0.1964 \\
$\overline \cos ({\bm a}_E, {\bm a}_C)$ & -0.6573 & -0.5945 & -0.5563 & -0.4561 & -0.3538 \\
$\overline \cos ({\bm a}_L, {\bm a}_P)$ & -0.9110 & -0.8761 & -0.8157 & -0.5273 & -0.2659 \\
\end{tabular}
\end{ruledtabular}
\end{table*}

\begin{table*}  % table 6
\caption{Pearson product-moment correlation coefficient $r$ for the scale-dependent Lagrangian and Eulerian accelerations
         at nondimensional time $St=10$. The correlation coefficient is determined for all three components
         of the accelerations. }
\label{tab_pearson}
\begin{ruledtabular}
\begin{tabular}{lrrrrrrrr}
$Ri$            &     0   &   0.1  &   0.2  &   0.5  &    1   \\ % &    2   &    5   &   10   \\
\hline
%$r$ (total)     &  0.0284 & 0.0510 & 0.0882 & 0.2852 & 0.6634 & 0.8443 & 0.9067 & 0.9371 \\
$r$ ($j=0$)     &  0.8584 & 0.9716 & 0.9892 & 0.9969 & 0.9995 \\ % & 0.9994 & 0.9997 & 0.9998 \\
$r$ ($j=1$)     &  0.8065 & 0.9347 & 0.9625 & 0.9957 & 0.9990 \\ % & 0.9996 & 0.9996 & 0.9998 \\
$r$ ($j=2$)     &  0.5208 & 0.6709 & 0.8287 & 0.9728 & 0.9921 \\ % & 0.9956 & 0.9971 & 0.9982 \\
$r$ ($j=3$)     &  0.1988 & 0.3427 & 0.5072 & 0.8257 & 0.9512 \\ % & 0.9732 & 0.9848 & 0.9896 \\
$r$ ($j=4$)     &  0.0884 & 0.1380 & 0.2141 & 0.4959 & 0.7765 \\ % & 0.8684 & 0.9154 & 0.9410 \\
$r$ ($j=5$)     &  0.0308 & 0.0478 & 0.0739 & 0.2002 & 0.4528 \\ % & 0.5946 & 0.6614 & 0.7274 \\
$r$ ($j=6$)     &  0.0116 & 0.0196 & 0.0304 & 0.0993 & 0.2919 \\ % & 0.4174 & 0.4425 & 0.4690 \\
$r$ ($j=7$)     &  0.0037 & 0.0097 & 0.0248 & 0.1248 & 0.3337 \\ % & 0.4317 & 0.4189 & 0.4136 \\
$r$ ($j=8$)     & -0.0052 & 0.0200 & 0.0261 & 0.1707 & 0.3837 \\ % & 0.4652 & 0.4309 & 0.4151 \\
\end{tabular}
\end{ruledtabular}
\end{table*}

\begin{table*}  % table 7
\caption{Scale-dependent Lagrangian and Eulerian acceleration statistics for $Ri=0.1$ at nondimensional time $St=10$ showing the rms of the Lagrangian and Eulerian acceleration, $a_E$ and $a_L$, and the flatness, $Fl_{a_E}$ and $Fl_{a_L}$ for the total and the scale-dependent contributions at scale $2^{-j}$.}
\label{tab_laea_0010}
\begin{ruledtabular}
\begin{tabular}{lrrrrrrrrrr}
$j$              & total  &    0   &    1   &    2   &    3   &    4   &    5   &    6   &   7     &   8     \\
\hline
$a_L$            &  9.833 &  0.156 &  0.361 &  0.769 &  1.787 &  3.672 &  5.666 &  5.773 &   3.491 &   1.280 \\
$Fl_{a_L}$       & 26.041 &  4.933 &  3.468 &  4.706 &  4.968 &  6.844 & 10.979 & 39.138 & 119.480 & 252.821 \\
$a_E$            & 11.689 &  0.150 &  0.360 &  0.682 &  1.589 &  3.477 &  6.026 &  7.137 &   5.344 &  2.364  \\
$Fl_{a_E}$       & 13.413 &  4.296 &  3.304 &  4.103 &  5.061 &  5.981 &  8.114 & 12.607 &  27.600 & 65.829  \\
\end{tabular}
\end{ruledtabular}
\end{table*}

\begin{table*}  % table 8
\caption{Scale-dependent Lagrangian and Eulerian acceleration statistics for $Ri=1$ at nondimensional time $St=10$ showing the rms of the Lagrangian and Eulerian acceleration, $a_E$ and $a_L$, and the flatness, $Fl_{a_E}$ and $Fl_{a_L}$.}
\label{tab_laea_0100}
\begin{ruledtabular}
\begin{tabular}{lrrrrrrrrrr}
$j$              & total  &    0   &    1   &    2   &    3   &    4   &    5   &    6   &    7   &    8   \\
\hline
$a_L$            &  0.942 &  0.079 &  0.257 &  0.328 &  0.405 &  0.450 &  0.435 &  0.336 &  0.190 &  0.058 \\
$Fl_{a_L}$       &  4.111 &  5.690 &  3.210 &  4.042 &  4.064 &  4.248 &  6.316 & 11.850 & 15.906 & 38.999 \\
$a_E$            &  1.075 &  0.079 &  0.257 &  0.330 &  0.410 &  0.473 &  0.530 &  0.483 &  0.254 &  0.062 \\
$Fl_{a_E}$       &  5.691 &  5.700 &  3.210 &  4.021 &  4.034 &  4.166 &  6.524 & 10.411 & 21.338 & 66.034 \\
\end{tabular}
\end{ruledtabular}
\end{table*}

\end{document}